\documentclass[preprint2]{aastex} 
\usepackage{graphicx}
\usepackage{pdfpages}
\usepackage{natbib}
\usepackage{epstopdf}

\shorttitle{Progenitor Bias in Compact Red Galaxies}

\begin{document}  

\bibliographystyle{apj}

\title{Evidence for (and Against) Progenitor Bias in the Size Growth of Compact Red Galaxies}

\author{Stephanie K. Keating\altaffilmark{1}, Roberto G. Abraham\altaffilmark{1}, Ricardo P. Schiavon\altaffilmark{2}$^{,}$\altaffilmark{3}, Genevieve Graves\altaffilmark{4}$^{,}$\altaffilmark{5}, Ivana Damjanov\altaffilmark{6}, Renbin Yan\altaffilmark{7}, Jeffrey Newman\altaffilmark{8}, Luc Simard\altaffilmark{9}}

\altaffiltext{1}
{Department of Astronomy and Astrophysics, University of Toronto,
50 St. George Street, Toronto, ON, M5S~3H4, Canada}

\altaffiltext{2}
{Gemini Observatory, 670 North A'ohoku Place, Hilo, HI 96720, USA}

\altaffiltext{3}
{Astrophysics Research Institute, Liverpool John Moores University, 146 Brownlow Hill, Liverpool L3~5RF, UK}

\altaffiltext{4}
{Department of Astronomy, University of California, Berkeley, CA 94720, USA}
\altaffiltext{5}
{Department of Astrophysical Sciences, Princeton University, Princeton, NJ 08544, USA}

\altaffiltext{6}
{Harvard-Smithsonian Center for Astrophysics, 60 Garden Street, Cambridge, MA 02138, USA}

\altaffiltext{7}
{Department of Physics and Astronomy, University of Kentucky, 505 Rose Street, Lexington, KY 40506-0055, USA}

\altaffiltext{8}
{Department of Physics and Astronomy, University of Pittsburgh, 3941 O'Hara Street, Pittsburgh, PA 15260, USA}

\altaffiltext{9}
{National Research Council of Canada, 5071 West Saanich Road, Victoria, BC V9E~2E7, Canada}


\begin{abstract}
Most massive passive galaxies are compact at high redshifts,  but similarly compact massive galaxies are rare in the local universe.  The most common interpretation of this phenomenon is that massive galaxies have grown in size by a factor of  about five since redshift $z=2$. An alternative explanation is that recently quenched massive galaxies are larger (a ``progenitor bias").  In this paper we explore the importance of progenitor bias by looking for systematic differences in the stellar populations of compact early-type galaxies in the DEEP2 survey as a function of size. Our analysis is based on applying the statistical technique of bootstrap resampling to constrain differences in the median ages of our samples and to begin to characterize the distribution of stellar populations in our co-added spectra. The light-weighted ages of compact  early-type galaxies at redshifts $0.5 < z < 1.4$ are compared to those of a control sample of larger galaxies at similar redshifts. We find that massive compact early-type galaxies selected on the basis of red color and high bulge-to-total ratio are younger than similarly selected larger galaxies, suggesting that size growth in these objects is {\em not} driven mainly by progenitor bias, and that individual galaxies grow as their stellar populations age. However, {\em compact early-type galaxies selected on the basis of image smoothness and high bulge-to-total ratio are older than a control sample of larger galaxies}.  Progenitor bias will play a significant role in defining the apparent size changes of early-type galaxies if they are selected on the basis of the smoothness of their light distributions.  
\end{abstract}

\section{Introduction}
\subsection{Context}
\label{sec:context}

One of the most surprising recent developments in galaxy evolution has been the discovery of a population of massive compact quiescent galaxies (``red nuggets") at high redshifts. These objects, first reported by \cite{Daddi2005}, have been the subject of over 40 observational papers. A representative subset of these would include \cite{Longhetti2007, Trujillo2007, Toft2007, Zirm2007, Cimatti2008, vanDokkum2008, Buitrago2008, Damjanov2009, Newman2010, Szomoru2010, vanDokkum2010, Mancini2010, Damjanov2011, Bruce2012, Szomoru2012, Law2012, Ryan2012, McLure2013,  Chang2013, Barro2013, ANewman2012, Patel2013}. These many papers explore in detail the potential sources of systematic error which could cause  sizes to be underestimated, or masses to be overestimated, or some combination of both, due to errors in photometric redshifts, errors in conversion from light to stellar mass, undetected extended envelopes that cannot be observed because of cosmological dimming, and other factors. Having been through this crucible, there is now broad consensus that massive quiescent galaxies at high redshifts are a factor of two to five smaller than local systems at similar mass.

In the present paper we treat red nuggets as an observationally established phenomenon, though it is important to note that two studies do disagree with this characterization. In the first study, \cite{Valentinuzzi2010} report little evidence for a changing fraction of very compact galaxies in rich clusters from $z=0.7$ to $z=0$.  This result might be understood as an environmental effect, although the role of environment is controversial \footnote{\cite{Raichoor2012} investigated the mass-size relation at $z \sim 1.2$ for morphologically selected early-type galaxies in field, cluster, and group environments, and found that for masses $10 < $log$(M/M_{\Sun}) < 11.5$, field galaxies appear to be larger than cluster galaxies at fixed stellar mass.  However, using DEEP3 at lower redshift but the same stellar mass range, \cite{Cooper2012} find the opposite trend: cluster galaxies appeared larger.  Using CANDELS data, \cite{Papovich2012} also find larger galaxies in the cluster environment at $z = 1.62$.  \cite{Zirm2012} find that passive galaxies in a proto-cluster at $z \sim 2$ are larger than their field counterparts.  Furthermore, studies by \cite{Maltby2010} and \cite{Rettura2010} find no trend with environment at $z < 0.4$ and $z \sim 1.2$ respectively. Recently, looking at galaxies in the COSMOS survey, \cite{HuertasCompany2013} found that the galaxy size-mass relation and size growth do not depend on environment.}.
 \cite{Saracco2010} is the second study that disputes the observational evidence for red nuggets: they report little change in the number density of compact quiescent galaxies from $z=1.5$ to $z=0$.  This investigation remains an outlier. 

Most attempts to understand the nature of red nuggets have assumed that they have some connection to local elliptical galaxies. In the last several years, new studies have begun to challenge this basic assumption \citep{vanderWel2011, Chevance2012, Bruce2012, Patel2013}. It has become clear that the structure of high-redshift quiescent compact galaxies does not resemble that of local elliptical galaxies. While the ellipticity distribution of nuggets resembles that of massive local spheroids, their S\'ersic indices are better matched to those of massive local disks.  The incompatibility between the bivariate ellipticity - S\'ersic index distribution of nuggets and any homogeneous local population \citep{Chevance2012}  means that the morphology of nuggets is presently a mystery. They may be a population of early-type galaxies with intrinsic shapes that differ from their local counterparts, or they may be disks with unusually massive bulges, or they may be a mix of these. They may even be a new class of galaxies unique unto themselves. An intriguing suggestion, proposed in the model of \cite{Hopkins2009b}, is that they may be the dense central component of early-type galaxies, which recent observations suggest may be better described by multi-component models \citep{Huang2013b, Huang2013}. 

Most authors have assumed that the evolving sizes seen in the red nugget population are
due to the physical expansion of individual galaxies\footnote{Although as early as 2008 van Dokkum et al. noted that galaxies forming (and ultimately joining the red sequence) at later times may be systematically
larger because they are less gas rich. We will return to this idea below.}.
Various mechanisms have been proposed to explain this expansion, such as mergers \citep[e.g.][]{Khochfar2006, Naab2007, Hopkins2009, Bezanson2009} or adiabatic expansion caused by extreme mass loss -- perhaps caused by quasar feedback or stellar winds \citep[e.g.][]{Fan2008, Damjanov2009, Fan2010}.  These papers all point to relevant physics that can contribute to size growth, but no model is completely satisfactory. The currently favored model is one where most of the growth comes from minor gas-poor (dry) mergers \citep[Lopez-Sanjuan et al. 2012]{Hopkins2010b, Naab2012, Trujillo2012, McLure2013}. However, the number of mergers required to explain the size evolution is much larger than what is predicted by $\Lambda$CDM models, which creates many more massive galaxies than are seen in the local universe \citep{Saracco2011}.
\cite{Shankar2010} used semi-analytic models based on a hierarchical growth of galaxies that are driven by an initial major, wet merger and followed by a number of late, minor, dry mergers and showed that compact galaxies at high-$z$ can grow on to the same local size-age relation; however, the models provide a poor match to the local size-mass relation.
Perhaps the greatest challenge to the idea that dry minor mergers alone can explain the observed size growth has come from \cite{ANewman2012}, who have used
very deep CANDELS data to demonstrate that there are simply not enough companions around high-$z$ galaxies
to account  for the very rapid size growth seen from $z=2.5$ to $z=1$. It seems that a two-phase mechanism is needed in which rapid early size growth is later augmented by a more gradual growth from minor mergers \citep{Oser2010,Oser2012}.

Returning  to observations, we have already noted that the many obvious sources of systematic error have been ruled out as the explanation for the observed size growth of massive  galaxies. However another source of concern is the possibility that the abundance of local analogs to the high-redshift nuggets may have be greatly underestimated. Initial studies based on data from the Sloan Digital Sky Survey (SDSS) have indicated an almost complete absence of very compact massive systems nearby \citep{Trujillo2009, Taylor2010} lending credence to the idea that the local galaxy size-mass relation is the result of a significant amount of size evolution on the part of red nuggets. Using the Wide-field Nearby Galaxy-clusters Survey, \cite{Valentinuzzi2010b} found that in nearby ($z \sim 0.05$) galaxy clusters, superdense galaxies represent nearly 22 \% of all cluster members with stellar mass range $3 \times 10^{10} \le M_{*/M_{\Sun}} \le 4 \times 10^{11}$, and have masses and sizes similar to their high-$z$ counterparts. 
\cite{Poggianti2013} looked for field superdense galaxies at $z=0.03-0.11$ using the Padova-Millennium Galaxy and Group Catalogue (PM2GC) and found that compact galaxies with radii and mass densities comparable to high-$z$ massive, passive galaxies represent 4.4\% of all galaxies with stellar masses $> 3 \times 10^{10} M_{\Sun}$, and claim that when stellar age and environmental effects are accounted for, the size evolution of galaxies between high and low $z$ is only a factor of $\sim 1.6$.  \cite{Poggianti2013b} also compare the number densities of compact galaxies at low redshift (using PM2GC) with the high redshift CANDELS results from \cite{Barro2013} and find little difference. Finally, in a recent study, \cite{Damjanov2013} identify nine compact, quiescent galaxies from SDSS with dynamical masses $M_{\mathrm{dyn}} > 10^{10} M_{\Sun}$, initially classed photometric point sources, but with redshifts $0.2 < z < 0.6$.
  If the abundance of local nuggets is greatly underestimated then this opens the door to ``progenitor bias" being the dominant
source of the observed size growth. The central idea here is that  
galaxies forming at later times (and ultimately joining the red sequence) may be systematically
larger because they are less gas rich \citep{vanDokkum2008}. Gas-rich systems forming earlier are 
losing total energy through dissipative processes while conserving mass, so 
the final galaxy is more compact.

The importance of progenitor bias  has been explored in detail in a number of papers, with most authors concluding that it is unlikely to be the dominant effect \citep[e.g.][]{vanDokkum2008, vanderWel2009, Hopkins2009b, Szomoru2011, Whitaker2012, LopezSanjuan2012, Bruce2012}. However, \cite{Carollo2013} have recently suggested that many compact galaxies may be missing from local catalogs due to misclassification as stars and/or intra-sample inconsistency in the definition of compactness.  
These authors use data from the COSMOS survey to argue that progenitor bias is the dominant source of observed size growth, noting that
 the {\em `apparent disappearance of Q-ETGs [quenched early-type galaxies] at later epochs may thus be a false reading of a reality in which earlier populations of denser Q-ETGs remain relatively stable in terms of numbers through cosmic time, but become less and less important, in relative number, at later and later epochs'.}

The central prediction of progenitor bias is that younger galaxies are larger at a fixed mass. At low redshifts there appears to be some support for this prediction \citep{Shankar2009, vanderWel2009}. Interestingly, there appears to be considerable morphological dependence in the age-mass relation: \cite{Bernardi2010} find little evidence for age-dependent sizes at fixed mass for elliptical galaxies, but show that large S0 and Sa galaxies tend to be younger at a fixed dynamical mass, suggesting that progenitor bias might be more important for early-type systems with disks. In the phenomenological picture of \cite{Huang2013,Huang2013b}, the innermost component of massive, early-type galaxies has a low S\'ersic index. Furthermore, at high redshift, the disk fraction of compact galaxies appears to be over 50\%  \cite[e.g.][]{vanderWel2011}. This suggests that the early phase of the nugget phenomenon is associated with disk galaxies, which appears to be consistent with the suggestion that many of the nuggets are indeed disks \citep[e.g.][]{vanderWel2011, Chevance2012, Bruce2012, Patel2013}.

On the other hand, at higher redshifts there appears to be no evidence for age-dependent galaxy sizes at fixed mass \citep{Trujillo2011, Whitaker2012}.  Perhaps this is because outside the local universe it is difficult to tell the difference between sub-classes of early-type galaxies. Existing investigations make no attempt to distinguish elliptical galaxies from the S0/Sa-like systems that may be an important component of the population of nuggets. It is therefore of considerable interest to look for trends in galaxy age as a function of redshift with
an eye toward understanding the importance of the morphological ``fine structure" 
used to distinguish elliptical galaxies from other types of objects in the early-type galaxy family  (e.g. S0 and Sa galaxies).

\subsection{Goal of This Paper}
In this paper, we aim to better characterize the link between stellar population age, galaxy size and morphology
as a function of redshift. 
Colors alone cannot be used to infer ages because of age-metallicity degeneracies and other factors described below.
On the other hand, the absorption features used to characterize galaxy ages on spectra are difficult to observe at the required signal-to-noise levels at high-redshifts (except in the cases of the most extreme post-starburst systems, \cite[e.g.][]{Bezanson2013, Whitaker2013}.  In this paper we describe an attempt to get around this basic difficulty
by using a new statistical technique we have developed for exploring galaxy ages using co-added spectra. By examining the light-weighted ages of galaxies as a function of size using this technique, our goal is to explore whether the red nugget phenomenon is more closely related to the physical expansion of galaxies already established on the red sequence, or whether it is likely due to  some form of progenitor bias. A measurement of younger ages for compact galaxies, as compared to larger galaxies, would indicate that the galaxy growth scenario is favored. A measurement that finds the opposite, that compact galaxies are older than larger galaxies, would suggest that progenitor bias is the preferred model.
We also seek to test whether the morphological ``fine structure" of galaxies (which we characterize crudely using apparent smoothness) leaves an imprint on their measured stellar populations. 

A plan for this paper is as follows.
Our methodology is described in \S\ref{sec:methodology}, with our subsamples defined in \S\ref{sec:samples}.  
Section \ref{sec:agemeasure} describes our age measurements, and \S\ref{sec:bootstrap} describes the bootstrap resampling performed on the galaxies that contribute to each co-added spectrum to generate a distribution of possible ages.  
Our results are discussed in \S3, and conclusions are presented in \S4.
Throughout this paper, we assume a $\Lambda$CDM cosmology with $H_{0} = 70$ km s$^{-1}$, $\Omega_{m} = 0.3$, and $\Omega_{\Lambda} = 0.7$.

\section{Methodology}
\label{sec:methodology}

\subsection{Concept}
\label{sec:concept}

The light-weighted age of a galaxy is often inferred from analysis of the strengths of age indicators, such as Balmer lines, in its integrated spectrum \citep[e.g.][]{Gonzalez1993,Worthey1994a,Terlevich2002, Schiavon2007, Trager2009}. If present, hot, young stars dominate the integrated light at UV-optical wavelengths, such that recent episodes of star formation skew the spectral energy distribution (SED) towards a young light-weighted age.\footnote{A simple example provided by \cite{Trager2008} shows that adding 2\% by mass of 1 Gyr old stars to a 12 Gyr old population results in skewing the apparent single-stellar-population-equivalent age (the age that an object would have if formed at a single time with a single chemical composition) of a galaxy to 5 Gyr (though, note that these numbers depend heavily on the age indicator adopted, as discussed in \cite{Schiavon2007}.)}.

Because it is extremely difficult to obtain spectra which have high enough signal-to-noise to measure an accurate light-weighted age, we chose to co-add spectra from selections of the DEEP Extragalactic Evolutionary Probe 2 (DEEP2) spectroscopic redshift survey \citep{Davis2003, Davis2005, Newman2012} in order to obtain a high-quality representative spectrum from which we can measure an age. 
Individual DEEP2 spectra have continua with an average signal-to-noise per pixel of $< 1$ \citep{Newman2012}, although many of the galaxies at the magnitudes probed in the present paper have signal-to-noise levels of $\sim 5 {\rm \AA}^{-1}$ (in their continua). This is clearly insufficient for accurate measurement of Lick indices, from which the light-weighted ages are determined.  
According to \cite{Graves2008}, in order to estimate ages with an error of $\approx 3$ Gyr, Lick indices should be measured with $\pm 20\%$ accuracy, which requires spectra with S/N $\sim 30-50 {\rm \AA}^{-1}$ \citep{Cardiel1998}, corresponding to $\sim 36-100$ DEEP2 spectra.

Galaxy populations are not perfectly homogeneous, and co-adding their spectra can be quite a risky exercise. As a cautionary example,
consider the outcome of co-adding spectra selected in a perfectly fair way from a parent population of galaxies whose age distribution is strongly bimodal. The co-added spectrum 
might (arguably) represent an adequate description of the summed properties of the stars within the galaxies,  but it would certainly
not resemble the intrinsic spectrum of any individual object. This suggests that careful attention must be given to pre-selection of objects prior to
co-addition, so that the objects being summed (to improve the total signal-to-noise) are intrinsically quite similar. This is clearly a potential source of bias. 
One novel aspect of the present paper
is a technique we have developed to better understand the underlying homogeneity of the sample
being co-added (see \S2.6). 

In spite of the risks, co-addition has been
used in attempts to measure the properties of similar objects in many areas of astronomy, 
particularly in cases where high signal-to-noise observations 
of individual objects are difficult to obtain\footnote{As early as 1985, \cite{Adelman1985} used co-addition to study the ultraviolet and optical region of a horizontal branch star in the field. Although instrumentation has improved drastically since then, observations of faint or distant objects still often benefit from co-addition.  For example, \cite{Dressler2004} utilized composite spectra to quantify general trends in star formation for galaxy populations at $z > 0.3$.   In order to investigate the distribution of metals in galaxies, \cite{Gallazzi2008} used co-added spectra of galaxies with similar velocity dispersions, absolute r-band magnitude and 4000 \AA-break values to probe areas of parameter space where their individual spectra had low signal-to-noise.}.  
In particular, co-addition has been used before to analyze the spectra of red galaxies. For example, \cite{Eisenstein2003} divided SDSS spectra of 22,000 luminous, red, bulge-dominated galaxies into subsamples selected on the basis of luminosity, environment, and redshift, then stacked them to make average spectra with high signal-to-noise ratios. These average spectra were found to be remarkably similar.  \cite{Schiavon2006} stacked DEEP2 spectra of red field galaxies with weak to no emission lines at $0.7 < z < 1$. They compared the stacked spectra to models of stellar population synthesis and showed that red galaxies at $z \sim 0.9$ have mean luminosity-weighted ages on the order of 1 Gyr and metallicities that are at least solar. Our procedure, outlined in \S\ref{sec:coadd} and \S\ref{sec:agemeasure}, follows very closely upon the methodology adopted in \cite{Schiavon2006}.
\cite{Cimatti2008} also used co-addition to study red nuggets at $z > 1.4$ by combining 13 early-type galaxy spectra from the GMASS project into a stacked spectrum with an equivalent integration time of $\approx\-\-500$ h.
The key to successful use of co-addition for our present purposes is to keep careful track of potential systematics and to ensure that our conclusions are based on comparison with a control sample that shares these systematics.

\subsection{Co-addition of DEEP2 Spectra}
\label{sec:coadd}

DEEP2 was designed to study galaxy evolution out to redshifts of $z \sim 1.4$ and targeted $>$ 50,000 galaxies over four widely separated fields covering a total sky area of 2.8 deg$^{2}$, observed with $\sim 1$ hr exposure times to a limiting apparent magnitude of $R_{\mathrm{AB}} < 24.1$.  The survey used the DEIMOS spectrograph on the 10-m Keck II telescope with the 1200-line mm$^{-1}$ grating which delivers high spectral resolution of $R \sim 6000$ with an observed wavelength range of $6500 < \lambda [\mathrm{\AA}] < 9200$  \citep{Faber2003} on the 10 m Keck II telescope.    Our galaxy samples are drawn from a subset of the DEEP2 sample known as the Extended Groth Strip (EGS: $\alpha=14^{h}17^{m}, \delta=+52^\circ 30'$). This field is the subject of a panchromatic study: the All-Wavelength Extended Groth Strip International Survey (AEGIS) \citep{Davis2007}, which includes \textit{Hubble Space Telescope} (\textit{HST})/Advanced Camera for Surveys (ACS) imaging.  Details of the DEEP2 observations, catalog construction, and data reduction can be found in \cite{Davis2003, Davis2005, Davis2007} and \cite{Newman2012}.

The DEEP2/AEGIS survey targeted $\sim$ 17,600 galaxies for spectroscopy. We use a subset of 2305 galaxies that contains redshifts, rest-frame $U$ and $B$ magnitudes, stellar masses and error on stellar masses, and (publicly available) quantified morphologies \citep{Cheung2012}. Structural parameters of the \textit{HST}/ACS images were measured using GIM2D, a two-dimensional bulge + disk decomposition program \citep{Simard2002}, providing bulge radii for S\'ersic indices $n=4$ and $n=2$, and bulge-to-total (B/T) ratios in $I$ and $V$ band for S\'ersic indices $n=4$ and $n=2$.  Stellar masses for the sample were derived by \cite{Bundy2006}: using \textit{BRIK} colors and spectroscopic redshifts, they fit the observed galaxy SEDs to a grid of model templates from \cite{Bruzual2003} with a \cite{Chabrier2003} initial mass function, spanning a range of star formation histories, ages, metallicities, and dust content.

We examined several choices of cuts to narrow down this catalog and select subsamples of the red, early-type galaxies of interest (by a combination of color, bulge-to-total ratio, and image smoothness; the exact properties selected are detailed in \S\ref{sec:samples}).  We used the $I$-band, $n=4$ cases for our selection of B/T values and bulge radii. 

Proprietary flux calibration algorithms, kindly provided by Renbin Yan, were applied to the one-dimensional spectra.  Each spectrum was  converted to rest-frame wavelengths and normalized by the mean flux at $4130 < \lambda [\mathrm{\AA}] < 4160$ (a range chosen in order to avoid relevant Lick indices).  Spectra that lacked data in the normalization range were excluded from further analysis and did not contribute to the co-addition.  The spectra were linearly interpolated onto a 0.5 \AA ~ grid.  The associated inverse variance (noise) spectra were normalized by the same factor as the galaxy spectra, and interpolated to the same grid.  The spectra were then co-added with each pixel weighted according to the inverse variance at that point: $ \frac{\sum_{i} y_{i}/\sigma_{i}^{2}}{\sum_{i} 1/\sigma_{i}^{2}} $.  Initial inspection of the co-added spectra revealed apparent contamination of H$\beta$ by the O2 $A$-band, which falls on the H$\beta$ line for redshifts $z \sim 0.56$.  In order to eliminate this (and other) contamination, both the O2 $A$- and $B$-bands (7594-7621 \AA ~ and 6867-6884 \AA ~ respectively) were masked from the rest-frame spectra by setting the inverse variances for those points to zero. By nature, inverse-variance-weighted co-addition gives a higher weight to brighter objects. This effect is reflected in the error estimates (see \S\ref{sec:bootstrap} and Figures \ref{fig:RBDhist} and \ref{fig:RSBDhist} for further details); we note that other stacking methods may yield tighter results.

In order to understand the systematics of our co-addition, for each point in the coadd we tracked the signal-to-noise spectrum (the signal-to-noise at a given point is defined as $\frac{f_{co}}{\sigma_{co}}$, where $f$ is the flux of the co-added spectrum and $\sigma_{co}$ is the standard deviation of the co-add, $\sigma_{co} = \frac{1}{\sqrt{\sum_{i} \frac{1}{\sigma_{i}^{2}}}}$), the mean redshift, the mean stellar mass, and the number of galaxies that contribute to the co-add.  Because we are primarily performing differential measurements between our samples, the systematics that co-addition is susceptible to can be minimized by realizing that whichever systematic affects a given co-added sample is similarly affecting the comparative co-added control sample.

\subsection{Sample Definitions}
\label{sec:samples}
The parent sample for our investigation consists of 2305 galaxies for which we had quantitative morphologies. The next challenge is to define a set of early-type galaxies from this parent sample. As shown by \cite{Moresco2013}, the observed properties of early-type galaxies are highly dependent on the way in which those galaxies are defined.  

We therefore examined two different samples of `early-type' galaxies, both of which are based on reasonable assumptions about the expected properties of these systems. These cuts are detailed in Table 1 and are described as follows:  

\begin{enumerate}
\item{\em Red \& Bulge-Dominated (RBD)}: 
We based this selection on cuts for rest-frame color ($U-B$ $> 0.9$) to select the red galaxies, and bulge-to-total ratio B/T $> 0.5$ to eliminate galaxies with disks.  Before implementing a mass cut, this sample is comprised of 203 galaxies.

\item{\em Red, Smooth \& Bulge-Dominated (RSBD)}:
\cite{Simard2002, Simard2009} suggests that usage of a measure of image ``smoothness" can aid in maximizing the number of E/S0 galaxies selected, while minimizing contamination from Sa-Irr type galaxies, by removing those with clumpy structure.  The image smoothness is defined as $S = R_{T} + R_{A}$, where $R_{T}$ and $R_{A}$ are indices that quantify the amount of light in symmetric and asymmetric residuals (respectively) from a fitting model, expressed as a fraction of the total galaxy model flux.  They are defined in detail in \cite{Simard2002}.  \cite{Simard2009} find that the optimal definition of an early-type galaxy is one with a limit of smoothness, $S2$ ($S$ measured within two half-light radii) of $S2 < 0.075$ (measured in the I-band) and B/T $> 0.35$; we therefore adopted these criteria for our second selection, in addition to a rest-frame color cut of $U-B > 0.9$ to ensure that we are picking out purely red galaxies.  Before implementing a mass cut, this sample is comprised of 119 galaxies.
\end{enumerate}

Each of the samples outlined above was then further subdivided by mass and the $I$-band half-light radius of the bulge. The mass division was performed with the intention of separating out the effects of mass from our measurements.  At masses $M > 10^{11} M_{\Sun}$, there is negligible evolution of the stellar mass function from $z=1$ to $z=0$ \citep[e.g.][]{Fontana2004,Bundy2006,Borch2006,Vulcani2011}, therefore we divided our samples between ``heavy" ($M_{*} > 10^{11} M_{\Sun}$) and ``light"  ($M_{*} < 10^{11} M_{\Sun}$) subsamples.  We found that the lighter galaxies had a signal-to-noise that was insufficient for measuring ages with a reasonable degree of certainty; therefore, we consider only the galaxies with $M_{*} > 10^{11} M_{\Sun}$ in our analysis.  After this mass cut, the RBD sample is comprised of 87 galaxies, 78 of which have data in our normalization range of 4130-4160 \AA, and the RSBD sample is comprised of 72 galaxies, 69 of which have data in the normalization range.

An ideal radius cut would be on the order of 1 kpc, which is the typical measured size of a red nugget; however, implementing such a cut on our sample resulted in too few compact galaxies to contribute to a meaningful co-added spectrum. Given this constraint, we implemented a radius division at 2 kpc, separating the above groups into ``compact" ($r < 2$ kpc) and ``control" ($r > 2$ kpc) subsamples.  This slightly larger-than-ideal radius division is still keeping with the accepted definition of a ``compact" galaxy: for example, \cite{Cassata2011} present a distinction between ``ultra-compact," ``compact," and ``normal" early-type galaxies (ETGs) based on a galaxy's location relative to the local mass-size relation, with a ``normal" ETG having $r_{e} \sim 2 - 4$ kpc at $M \gtrsim 10^{10.6} - 10^{10.8} M_{\Sun}$. 

\begin{figure*}[hbt!]
\includegraphics[trim=18mm 10mm 10mm 0mm, angle=0, scale=0.8]{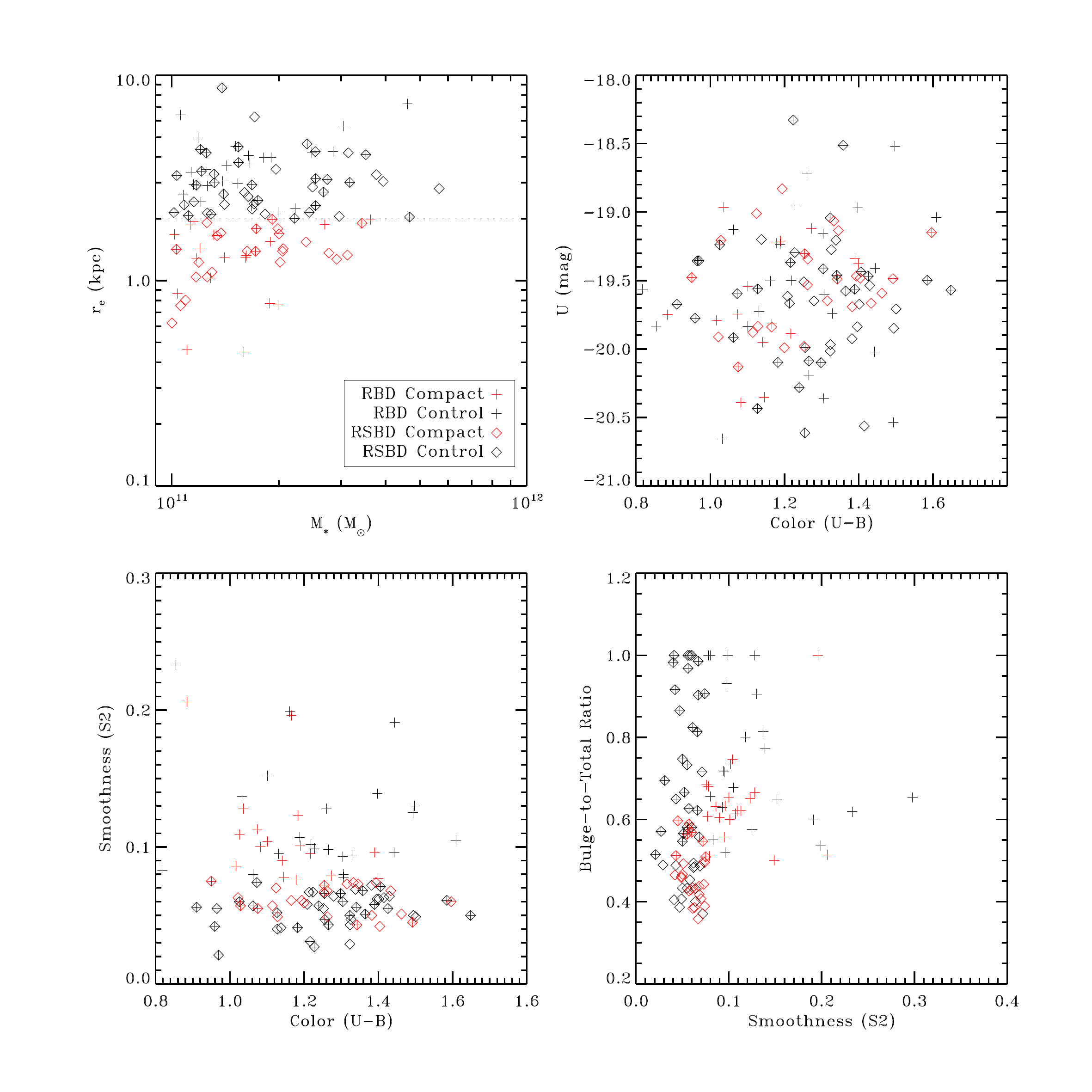}
\caption{The overlap of our samples in a mass-radius diagram (top left; dashed line indicates the radius division at $r_{e} = 2$ kpc, where $r_{e}$ is I-band half-light radius of the bulge), color-magnitude (top right), color-smoothness (bottom left), and smoothness-B/T (bottom right).  ``Red \& Bulge Dominated (RBD)" (B/T $>0.5$, $U-B > 0.9$, $M > 10^{11} M_{\Sun}$), marked by crosses (red for compact, black for control), and ``Red, Smooth \& Bulge Dominated (RSBD)" (B/T $>0.35$, S2 $< 0.075$, $U-B >0.9$, $M > 10^{11} M_{\Sun}$), marked by diamonds (red for compact, black for control).}
\label{fig:overlap}
\end{figure*}

Table 1 shows the number of galaxies remaining in our samples after each consecutive cut. Figure \ref{fig:overlap} shows the overlap of the two samples in mass-radius (with a dotted line indicating the radius division between the compact and control samples), color-magnitude, color-smoothness, color-B/T, and smoothness-B/T.


\begin{deluxetable}{ll}
\tablewidth{0pt}
\tabletypesize{\small}
\tablecaption{Galaxies Remaining After Consecutive Cuts}

\tablehead{
\colhead{Cut}		&	\colhead{\# Remaining}
} 
\startdata
Total					& 	2305\\
Remove bad radii		& 	2273\\
$U-B > 0.9$			&	839\\
$M > 10^{11} M_{\Sun}$ 	&	253 \\
\sidehead{RBD}
B/T  $>0.5$ 			&	87\\
$ r< 2$ kpc			&	30\\
$r > 2$ kpc			& 	57\\
\sidehead{RSBD}
B/T $> 0.35$			&	160\\
S2 $< 0.075$			&	72\\
$r < 2$ kpc			&	28\\
$r > 2$ kpc			& 	44\\

\enddata
\end{deluxetable}


\begin{figure*}
\includegraphics[trim=0mm 0mm 10mm 20mm, angle=0, scale=0.8]{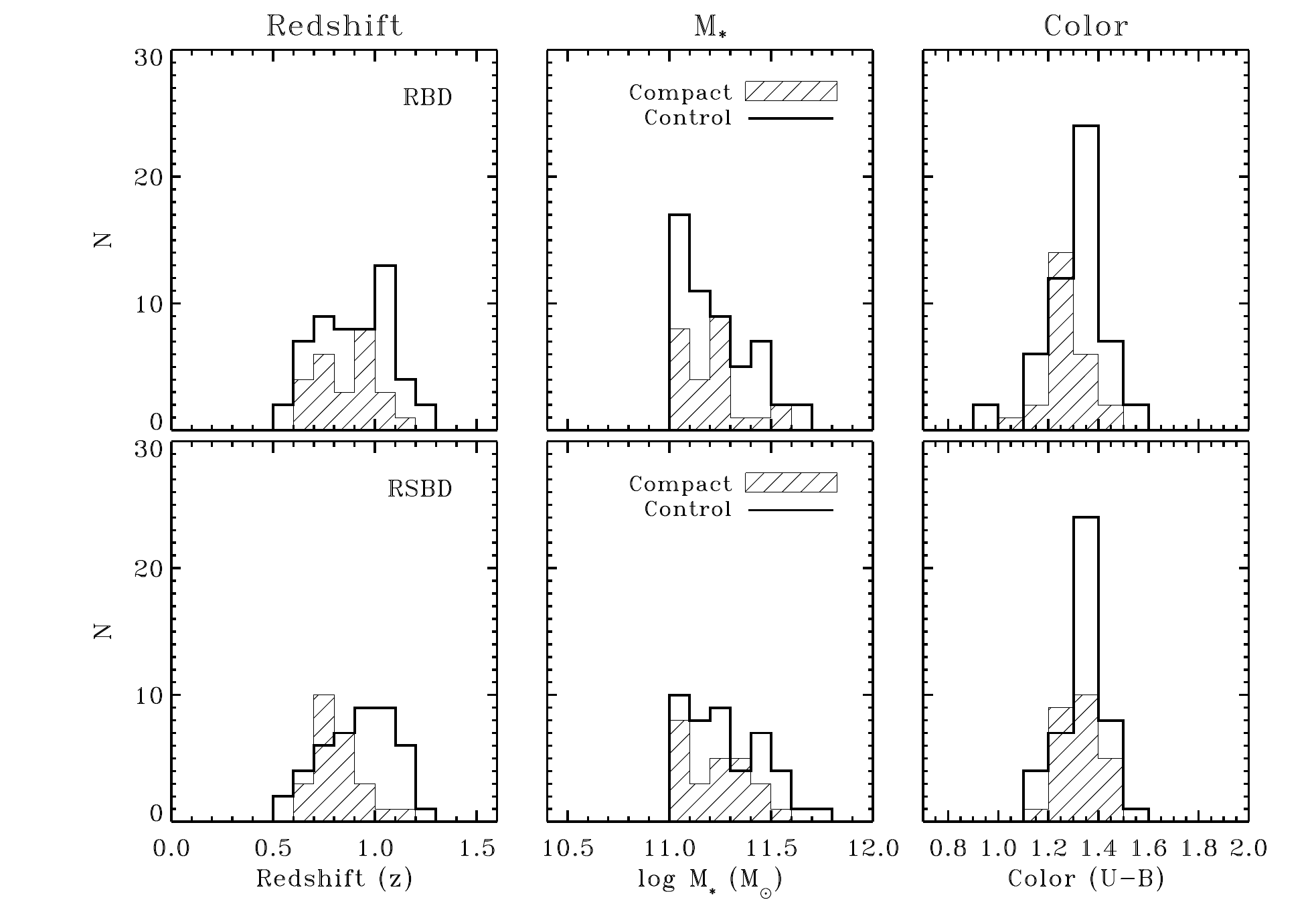}
\caption{Distributions in redshift (first column), stellar mass (second column), and color (third column) for the compact ($r_{e} < 2$ kpc) and control ($r_{e} > 2$ kpc) subsamples of the ``Red \& Bulge Dominated (RBD)" sample (top row), which implements cuts for B/T $>0.5$, $U-B > 0.9$, $M > 10^{11} M_{\Sun}$, and of the ``Red, Smooth \& Bulge Dominated (RSBD)" sample (bottom row), which implements cuts for B/T $>0.35$, S2 $< 0.075$, $U-B >0.9$, $M > 10^{11} M_{\Sun}$.  All histograms have a bin size of 0.1 in redshift/log($M_{*}$)/(U-B), respectively.}
\label{fig:c1hist}
\end{figure*}


\subsection{Systematic Properties of the Samples}
\label{sec:systematics}

The cuts to the DEEP2 catalogue made for each subsample are listed in Table 2, along with the number of galaxies in that subsample with data between 4130 and 4160 \AA (the range in which our spectra were normalized).  Figure \ref{fig:c1hist} shows the histograms of the distribution in redshift, stellar mass, and color for the subsamples defined in \S\ref{sec:samples}. We performed a two-sample Kolmogorov-Smirnov (K-S) test on the distributions in each panel of Figure \ref{fig:c1hist}. The K-S test reports a $p$-value, which is the probability that the two cumulative frequency distributions would be as far apart as they are measured if the two samples were randomly sampled from identical populations. The results of the K-S test indicated that the distributions in each panel are not statistically different: all have $p$-values of $p > 0.05$, with the exception of the RSBD redshift distributions, which had $p = 0.012 < 0.05$.

\begin{figure*}
\includegraphics[trim=0mm 0mm 0mm 0mm, angle=0, scale=0.8]{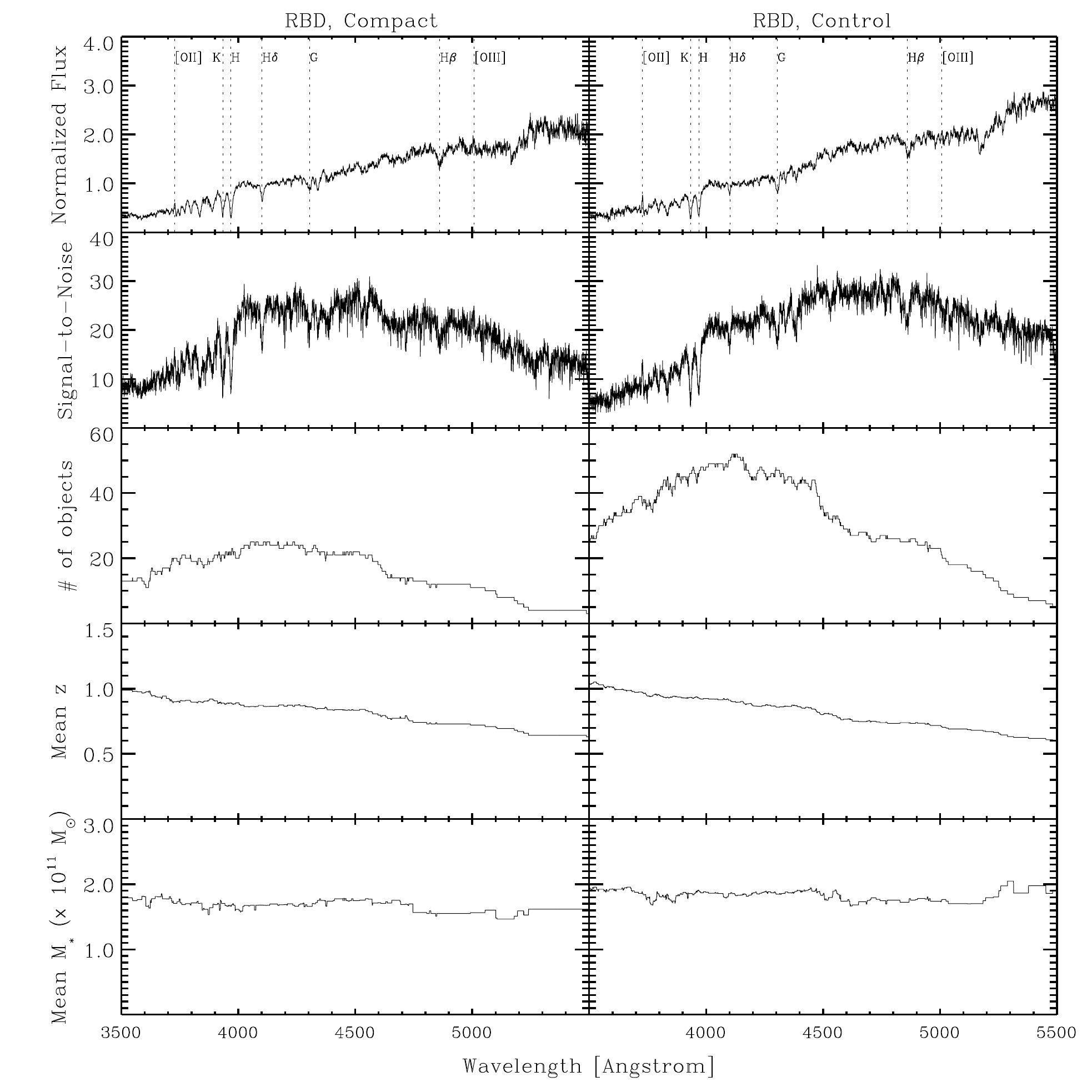}
\caption{The topmost panel shows the co-added spectrum for the compact ($r < 2$ kpc; left column) and control ($r > 2$ kpc; right column) galaxies for our ``Red \& Bulge Dominated (RBD) Sample" sample of galaxies (selected by color and bulge-to-total ratio).  Dotted lines mark the location of spectral features of interest (from left to right: [OII], Ca K and H, H$\delta$, G-band, H$\beta$, [OIII]).  The panels below display the tracked systematics, including signal-to-noise, the number of objects contributing at each given point in the co-addition, the average redshift, and the average stellar mass.}
\label{fig:c1spec}
\end{figure*}

\begin{figure*}
\includegraphics[trim=0mm 0mm 0mm 0mm, angle=0, scale=0.8]{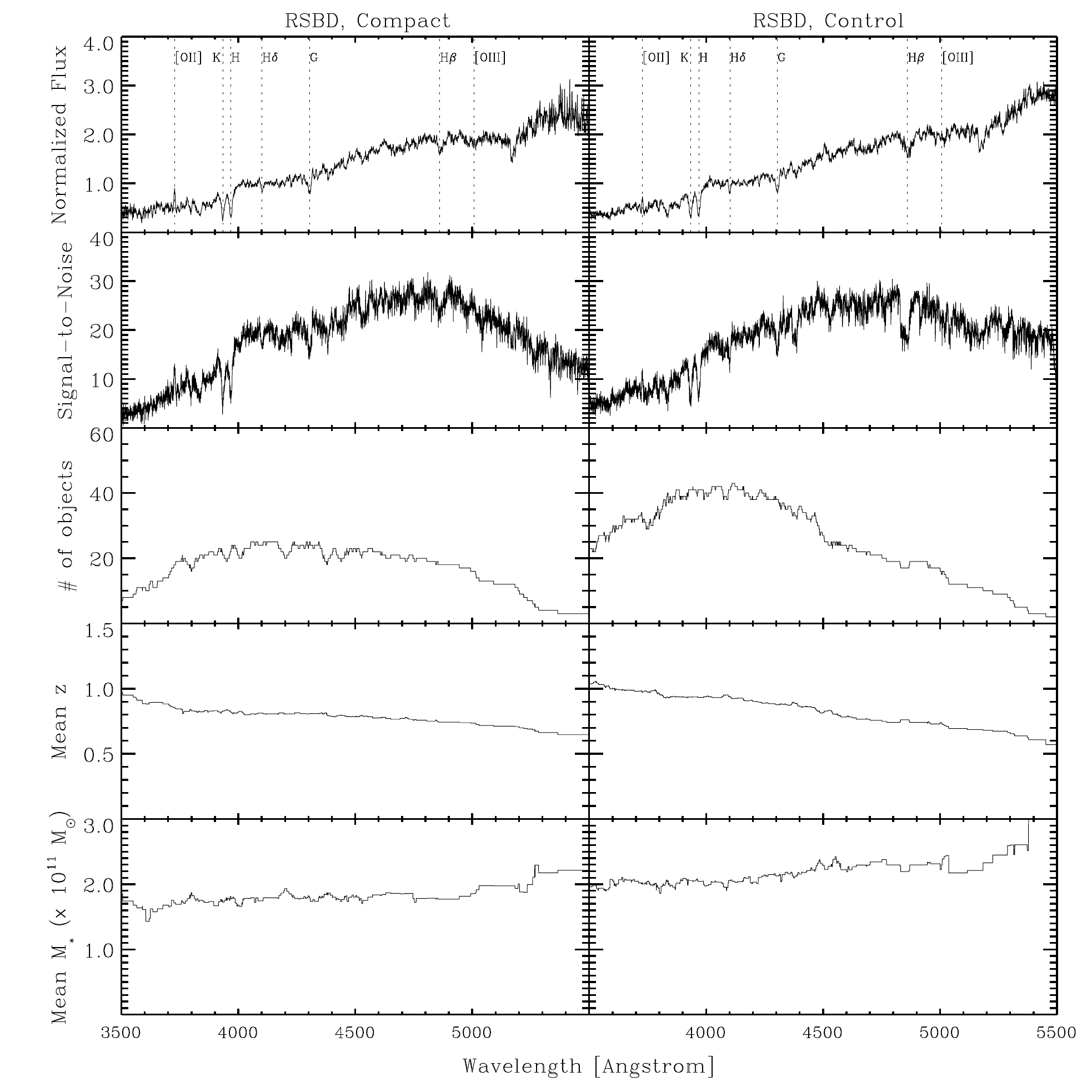}
\caption{The topmost panel shows the co-added spectrum for the compact ($r < 2$ kpc; left column) and control ($r > 2$ kpc; right column) galaxies for our ``Red, Smooth \& Bulge-Dominated (RSBD)" sample of galaxies (selected by image smoothness, bulge-to-total ratio, and color).  Dotted lines mark the location of spectral features of interest (from left to right: [OII], Ca K and H, H$\delta$, G-band, H$\beta$, [OIII]).  The panels below display the tracked systematics, including signal-to-noise, the number of objects contributing at each given point in the co-addition, the average redshift, and the average stellar mass.}
\label{fig:c3spec}
\end{figure*}

Figures \ref{fig:c1spec} and \ref{fig:c3spec} show the resultant co-added spectra and tracked systematics for the RBD Compact (left panel) and Control (right panel) subsamples and the RSBD Compact and Control subsamples, respectively.  The topmost panels show the co-added spectrum of the given subsample of galaxies.  Note the features of an early-type galaxy, such as the prominent 4000 \AA ~ break.  The spectra also feature prominent Balmer lines, which are strongest in the spectra of A-type stars, and indicative of stellar populations that are evolving as a result of a recent burst of star-formation, as well as Ca H- and K-lines, strongest in stars cooler than A-types.  We note that the RBD Compact sample displays markedly stronger Balmer lines than those of the other samples. The RBD Compact sample also shows signs of [OIII]-5007 emission\footnote{We do not have spatially resolved spectra for our sample, but the line fluxes are small.  While in principle extended emission line regions could bias our size estimates upward, and weak AGNs contamination could bias them downward, at the levels seen in our red sample neither effect will be large.}, which is characteristic of active galactic nuclei (AGNs) and/or young stellar populations \citep{Kauffmann2003}. The signal-to-noise ratio is highest where the number of objects contributing to the composite is highest, which occurs typically between $3500 < {\rm \AA} < 5500$.   For all co-added subsamples, the change in redshift follows a similar trend, decreasing from $z \sim 1$ to $z \sim 0.6$.  Both the compact and control galaxies also tend to have similar average masses.

\begin{deluxetable}{llllllclc}
\tablewidth{0pc}
\tabletypesize{\small}
\tablecaption{Details of Sample Cuts}

\tablehead{
\colhead{Sample}		&	\colhead{B/T}			&	\colhead{S2\tablenotemark{a}} 		& 	\colhead{$U-B$}		&  \colhead{Radius\tablenotemark{b}} 			& \colhead{Mass} 			& 	\colhead{No.\tablenotemark{c}}	& \colhead{S/N\tablenotemark{d}}	& \colhead{$\sigma$\tablenotemark{e}} \\

\colhead{Name}		&	\colhead{Cut}			&	\colhead{Cut} 		& 	\colhead{Cut}		&  \colhead{($R_{e}$) Cut}	& \colhead{Cut ($M_{\Sun}$)}	& 	\colhead{Remaining}		& \colhead{ }	 	& \colhead{(km s$^{-1}$)}
} 
\startdata
RBD, Compact			&	$>0.5$	&	N/A		&	$>0.9$			& $< 2$ kpc		& $> 10^{11}$	& 25		& 23.0	& 289	\\
RBD, Control			&	$>0.5$	&	N/A		&	$>0.9$			& $> 2$ kpc		& $> 10^{11}$	& 53		& 24.8	& 297	\\
RSBD, Compact		&	$>0.35$	&	$<0.075$	&	$>0.9$			& $< 2$ kpc		& $> 10^{11}$	& 25		& 23.0	& 299\\	
RSBD, Control			&	$>0.35$	&	$<0.075$	&	$>0.9$			& $> 2$ kpc		& $> 10^{11}$	& 44		& 22.4	& 301 \\
\enddata
 \tablecomments{Details of cuts made for each sample, along with other relevant information, including inferred velocity dispersions.}
\tablenotetext{a}{S2 characterizes image smoothness measured within two half-light radii. See \S\ref{sec:samples} for more information.}
\tablenotetext{b}{Where $r_{e}$ is $I$-band half-light radius of the bulge.}
\tablenotetext{c}{Number of galaxies with spectra in the normalization range of 4130-4160 \AA.}
\tablenotetext{d}{Signal-to-noise measured between 4000-5000 \AA.}
\tablenotetext{e}{See \S\ref{sec:agemeasure} for a description of the velocity dispersion measurement.}
\label{table:cuts}
\end{deluxetable}

\subsection{Age Measurements}
\label{sec:agemeasure}

To measure the light-weighted ages of the co-added spectra, we use the EZ\_Ages IDL code package \citep{Graves2008,Schiavon2007}, which computes the mean, light-weighted stellar population age for unresolved stellar populations, along with [Fe/H], and abundance ratios [Mg/Fe], [C/Fe], [N/Fe], and [Ca/Fe].  For convenience, a short description of the age-measurement process, including a brief description of the EZ\_Ages algorithm, is found below, but the reader interested in details is referred to \cite{Graves2008, Schiavon2007} for an extensive discussion of techniques used to measure light-weighted ages from faint galaxy spectra.

We measured Lick index line strengths for each coadded spectrum using the automated IDL code Lick\_EW, available as part of EZ\_Ages.  To measure the velocity dispersions, which are required for determining the Lick index line strengths, we used the cross-correlation method as implemented in the fv.fxcor routine in IRAF \citep{Tonry1979}.  The cross-correlation function (CCF) between the observed, co-added spectrum and a template spectrum of a single stellar population calculated for solar metallicity and age 2.0 Gyr (from the \cite{Schiavon2007} library of synthetic single stellar population spectra) was calculated.  The width of this CCF is sensitive to the widths of the absorption lines in the target spectrum (which are broadened by the velocity dispersion).  This sensitivity was used to estimate the velocity dispersion of the target \citep[e.g.][]{Schiavon2006}.  The template was convolved with a range of velocity dispersion values to create a library of sigma-broadened spectra, each of which was then cross-correlated with the original template spectrum.  We calculated the full width at half maximum (FWHM) of the CCF peaks and used this to determine the relationship between the FWHM and the velocity dispersion used to broaden the original template spectrum.  Next, the CCF  between co-added spectrum and the template spectrum was determined, and the FWHM of the CCF peak was measured.  This was input into the relation derived from the sigma-broadened spectra in order to obtain the velocity dispersion of the co-added spectrum.  The resultant velocity dispersions are listed in Table \ref{table:cuts}.

Next, the light-weighted age of each co-added spectrum was determined using EZ\_Ages.  Briefly, EZ\_Ages works by taking measured index strengths as input, and compares them to the stellar population synthesis models of \cite{Schiavon2007}.  If they are provided, EZ\_Ages uses errors in the Lick index data to estimate the uncertainties in the ages, [Fe/H], and abundance ratios, and uncertainties are assumed to be dominated by measurement errors in the line strengths.  In our case, we used the signal-to-noise of each co-added spectrum to determine the errors. The models provide a choice of solar-scaled or $\alpha$-enhanced (average [$\alpha$/Fe] = +0.42) isochrones.  

EZ\_Ages uses a Balmer line and an iron line to break the age-metallicity degeneracy.  The default choices for these lines are H$\beta$ for age-sensitivity and an average of Fe5270 and Fe5335 ($\langle$Fe$\rangle$)for [Fe/H] sensitivity, though other lines or combinations of lines can be specified by the user.  To ensure we used lines that would provide the most accurate determination of age, we constructed a synthetic galaxy spectrum with a simple stellar population of solar metallicity using BC03 and tested several combinations of Lick indices with EZ\_Ages to fit the age of the galaxy at 0.5, 1, 2, 3, 5, and 7 Gyr.  We performed a simple percent error test on the ages calculated from EZ\_Ages to compare it to the true age of the ideal SSP. We chose the index combination that gave the least average percent error, with the fewest dropped fits, which was a combination of H$\beta$, H$\gamma_{F}$, and H$\delta_{F}$ for the Balmer lines, and $\langle$Fe$\rangle$ for the iron lines. Table \ref{table:indices} lists the age as measured by EZ\_Ages and the percent difference from the true age of the an ideal simple stellar population.


\begin{deluxetable}{ccc}
\tablewidth{0pt}
\tabletypesize{\small}
\tablecaption{Percent Error for EZ\_Ages Indices}

\tablehead{
\colhead{SSP Age (Gyr)}		&	\colhead{Age from EZ\_Ages (Gyr)} 	& \colhead{\% Error}
} 
\startdata
0.5	&	0.58		&	16.6 \\
1.0	&	0.98		&	2.2 \\
2.0	& 	1.88		&	6.2\\
3.0	&	2.99		&	0.1\\
5.0	&	5.05		&	1.0\\
7.0	&	6.21		& 	11.2\\
\enddata
\tablecomments{The percent differences are indicative of the achieved accuracy in the ideal case of an SSP.}
\label{table:indices}
\end{deluxetable}

\subsubsection{Systematic Errors on the Age Measurements}
\label{sec:infill}

We investigated two potential sources of systematic error on our age measurements: the use of a single template to measure the velocity dispersions and emission line infill of Balmer lines. In principle, neither effect should be large (because convolution with a velocity dispersion is a second-order effect on measurements of equivalent width, and because emission line infill should be small for quiescent galaxies\footnote{Furthermore, the relatively weak [OII] emission in the spectra does not necessarily indicate star formation. The ratio between [OII] and H$\beta$ is similar to that seen in low-ionization nuclear emission-line regions (LINERs) or other LINER-like galaxies \citep{Yan2006,Yan2012}. The [OII] emission seen here is likely similar to the extended emission-line region commonly seen by IFU surveys such as SAURON and ATLAS3D in a large fraction of nearby early-type galaxies \citep{Sarzi2010}. Such regions are likely produced by photoionization from old but hot stars, such as post-AGB stars, rather than by star formation.}), but it is worth verifying this. 

\begin{figure*}[hbt!]
\includegraphics[trim= 20mm 0mm 0mm 0mm, angle=0, scale=0.63]{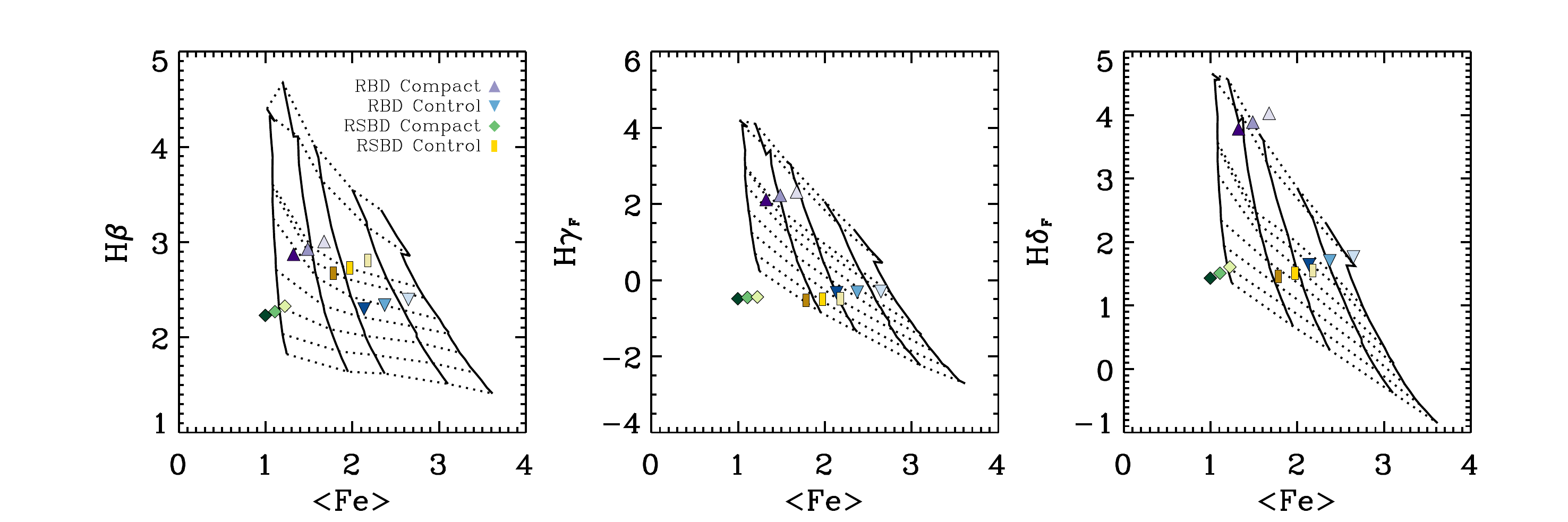}
\caption{Index-index model grids of H$\beta$ and $\langle$Fe$\rangle$ (left), H$\gamma_{F}$ and $\langle$Fe$\rangle$ (center), and H$\delta_{F}$ and $\langle$Fe$\rangle$ (right) showing the effects of changing the velocity dispersion by +20\% (lightest symbols) and -20\% (darkest symbols). The values at the measured velocity dispersion are denoted by the symbol colors in the legend. Solid lines show constant [Fe/H] from left to right of $-1.3$, $-0.7$, $-0.4$, $0.0$, and $+0.2$. Dotted lines show constant age from top to bottom of 1.2, 1.5, 2.5, 2.8, 3.5, 5.0, 7.0, 10.0, and 14.1 Gyr.}
\label{fig:sigma}
\end{figure*}

The effect of template mismatch (and/or velocity dispersion uncertainties) was estimated by altering the measured velocity dispersions by $\pm 10\%$ and $\pm 20\%$ and noting the resulting changes in the ages returned by EZ\_Ages. These changes are listed in Table \ref{table:sigma}, and Figure \ref{fig:sigma} shows how the velocity dispersion affects the samples on an index-index grid of three different Balmer lines and $\langle$Fe$\rangle$. With the exception of the RSBD Compact samples, ages change by around $\sim 0.2$ Gyr which makes no difference to our conclusions. The RSBD Compact sample fell off the model grids when the velocity dispersion was reduced by either 10 or 20\%, which is a clear sign that the velocity dispersions were too low. On the other hand, when the RSBD Compact velocity dispersion was increased the measured ages become older by $\sim 2$ Gyr, which makes the conclusions of the present paper (described below in \S\ref{sec:results}) even stronger. We note that in general, a higher velocity dispersion results in younger inferred ages, but the opposite is true for the RSBD Compact sample. The reason for this is evident in Figure \ref{fig:sigma}. At the measured velocity dispersion, the RSBD Compact sample is barely on the H$\beta$ vs. $\langle$Fe$\rangle$ grid, and is not on the grids of H$\gamma_{F}$ or H$\delta_{F}$ at all. When a measurement falls off the model grid for a particular index, that index is not used in the fit. When the velocity dispersion is increased (lightest symbols), the RSBD Compact sample falls on the model with age 7 Gyr for the H$\beta$ and the model with age 10 Gyr for H$\delta_{F}$. Because EZ\_Ages is weighted to take the average of the Balmer features, the resultant age is older for the RSBD Compact sample because the inferred age at the default velocity dispersion was determined only by the H$\beta$ index. The RSBD Control sample was better behaved, with a change of $\sim 0.4$ Gyr for $\sigma \pm 10\%$ and a maximum change of $+1.14$ Gyr for $-20 \%$. Our overall conclusion is that template mismatch/velocity dispersion uncertainties are unlikely to significantly impact our conclusions.

The amount of emission line infill of the Balmer absorption features was characterized by fitting stellar population models to the continua of the co-added spectra using routines adapted from SDSS analysis outlined in \cite{Brinchmann2004,Tremonti2004}. Briefly, the procedure is as follows: a library of template spectra was generated using \cite{Bruzual2003} stellar population synthesis code (BC03). The templates were composed of single stellar population models of 10 different ages (0.005, 0.025, 0.1, 0.2, 0.6, 0.9, 1.4, 2.5, 5.0, and 10.0 Gyr) and three different metallicities (20\%, 100\%, and 250\% $Z_{\Sun}$). The templates were convolved to the appropriate measured velocity dispersion of each co-added sample and re-sampled to match the data, and then a nonnegative least-squares fit was performed to construct the best-fitting model. Once the best-fitting stellar population has been subtracted from the continuum, any remaining residuals were removed, and the nebular features were fit. We then used the infill-corrected Balmer equivalent widths and measured the resulting changes in the ages returned by EZ\_Ages. These changes, along with the equivalent widths of the nebular features (negative values indicate emission features), are listed in Table \ref{table:infill}. 

The RBD sample ages change by $< 0.03$ Gyr.  The largest change shown is in the RSBD Control sample, which gets older by 0.68 Gyr. The RSBD Compact sample falls off the model grids and therefore does not have an age estimate. An inspection of the index-index grid of H$\beta$ and$\langle$Fe$\rangle$ for this sample reveals that it is just barely outside of the grid boundaries, and lies closest to the models with ages between 7 and 10 Gyr. This is consistent with the measured age of the original (non-infill-corrected) sample. Given these small changes, our conclusion is that the amount of infill does not vary enough across our samples to significantly impact the differences between our age measurements.


\begin{deluxetable}{lccccc}
\tablewidth{0pt}
\tabletypesize{\small}
\tablecaption{Systematics I: Effect on Age Measurements of Changing Velocity Dispersion}

\tablehead{
\colhead{Sample}	& \colhead{Age ($\sigma - 20\%$)}	 & \colhead{Age ($\sigma - 10\%$)}	&	\colhead{Nominal Age$^{\dagger}$} & \colhead{Age ($\sigma +10\%$)}	 & \colhead{Age ($\sigma +20\%$)} \\
\colhead{ }	& \colhead{(Gyr)} & \colhead{(Gyr)}	& \colhead{(Gyr)}	& \colhead{(Gyr)}	& \colhead{(Gyr)}
}
\startdata
RBD Compact		&	1.84	&	1.67 	&	1.50	&	1.33	&	1.11\\
RBD Control		&	3.43	&	3.00	&	2.78	&	2.43	&	2.92 \\
RSBD Compact	&	N/A	& 	N/A	&	7.27	&	9.40	&	9.03\\
RSBD Control		&	4.65	&	3.96	&	3.51	&	3.00	&	3.28\\
\enddata
\tablenotetext{\dagger}{The nominal age is the default age: measured on the original co-added spectrum, using the default index measurements as described in section \S \ref{sec:agemeasure} and the original measured velocity dispersions.}
\label{table:sigma}
\end{deluxetable}

\begin{deluxetable}{lccccc}
\tablewidth{0pt}
\tabletypesize{\small}
\tablecaption{Systematics II. Effect on Ages Measurements of Correcting Balmer Line Infill}

\tablehead{
\colhead{Sample}	& \colhead{H$\beta$ EW}		& \colhead{H$\delta_{F}$ EW} 	& \colhead{H$\gamma_{F}$ EW}	& \colhead{Nominal Age$^{\mathrm{a}}$} 	& \colhead{Infill} \\
\colhead{ }		& \colhead{(nebular)}		& \colhead{(nebular)}		& \colhead{(nebular)}			& \colhead{(Gyr)} 		& \colhead{Age (Gyr)}
}
\startdata
RBD Compact		& $0.069$		&	$-0.049$	&	$0.060$	&	1.50	&	1.48\\
RBD Control		& $-0.110$	&	$-0.011$	&	$0.081$	&	2.78	&	2.81\\
RSBD Compact	& $-0.062$	&	$0.051$	&	$-0.021$	&	7.27	&	N/A$^{\mathrm{b}}$\\
RSBD Control		& $-0.178$	&	$-0.109$	&	$-0.050$	&	3.51	&	4.19\\
\enddata
\tablenotetext{a}{The nominal age is the default age: measured on the original co-added spectrum, using the default index measurements as described in \S \ref{sec:agemeasure} and the original measured velocity dispersions.}
\tablenotetext{b}{The RSBD Compact sample fell off the model grids after correcting the Balmer features for emission line infill. An inspection of the index-index plot of $\langle$Fe$\rangle$ vs. H$\beta$ plot reveals that the sample is just barely outside of the grid boundaries, and is closest to the model with an age just older than 7 Gyr.}
\label{table:infill}
\end{deluxetable}


\subsection{Characterizing the Homogeneity of Co-added Spectra}

\label{sec:bootstrap}
Our approach to characterizing the age estimates inferred from co-added spectra
is based on the central ideas of ``bootstrap resampling" \citep{Efron1994}.
The statistical bootstrap
technique attempts to reconstruct the shape of an underlying distribution by 
resampling, with replacement, from observed data. This means that if the original data set has size $n$, a new, random sample of size $n$ is drawn from the original sample by allowing the same element to be drawn multiple times. A measurement is made from the new sample (in our case, the age of the co-added spectrum). This process is then repeated a large number of times.

The typical use of a bootstrap
is to place error estimates on observables, but the technique is actually 
more general than this, since the shape of the bootstrapped distribution
itself can also be used to probe the homogeneity of an underlying sample.  We recognize that the galaxy samples we have selected, although chosen to have similar properties, are nevertheless most likely not a homogeneous population.  Using the bootstrapping technique allows us to deal with this issue explicitly: if an underlying distribution is multi-modal, a record of this is traced by the bootstrap.

Our application of this useful aspect of the bootstrap is best illustrated using 
simulations. Using the \cite{Bruzual2003} stellar population synthesis code (BC03), we created several synthetic galaxy spectra as comparisons for our co-added DEEP2 spectra.  The synthetic spectra were produced by using simple stellar populations with 20\% solar metallicity ($Z=0.004$) at ages of 2 Gyr (we refer to this as the ``young" galaxy) and of 7 Gyr (the ``old" galaxy).  These spectra were convolved with a Gaussian to simulate a velocity dispersion of 300 km s$^{-1}$.  

\begin{figure*}[hbt!]
\includegraphics[trim= 20mm 10mm 0mm 0mm, angle=0, scale=0.8]{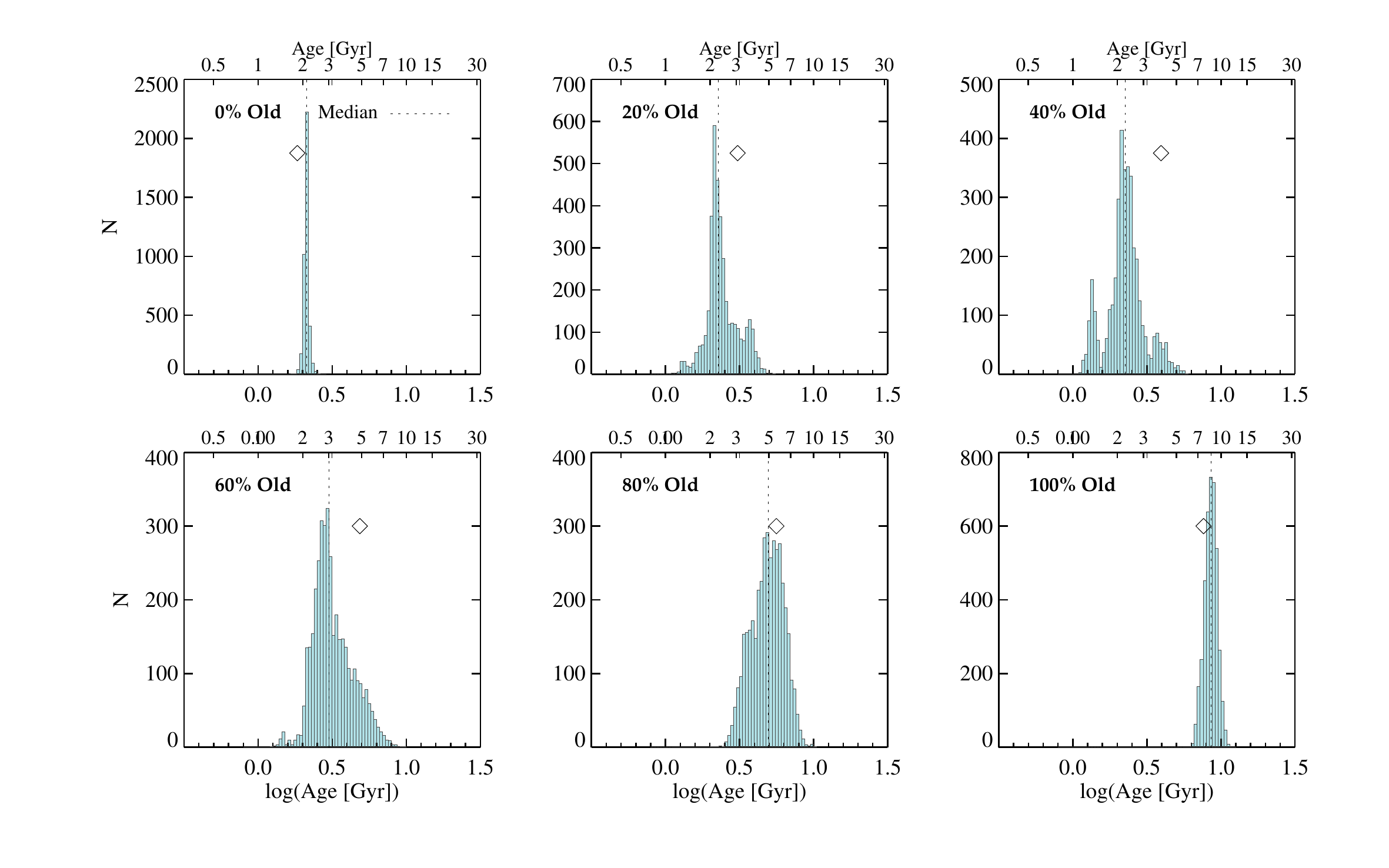}
\caption{Age histograms generated from 4000 bootstrap resamplings of the synthetic galaxy spectra drawn from parent populations with differing numbers of ``young" (2 Gyr) and ``old" (7 Gyr) spectra. The histograms have bin size 0.02 in log(age (Gyr)). The median log age is marked in each histogram by a dotted line. Black diamonds denote the nominal light-weighted age for the original co-added spectrum of each parent population. The homogeneous populations (0\% old and 100\% old) have Gaussian distributions, whereas the mixed populations have more skewed distributions.}
\label{fig:synthhisto}
\end{figure*}

We created six ``parent" galaxy populations comprised of 25 galaxies each, to mimic our smallest sample of galaxies. We added noise to each galaxy spectrum in the parent populations such that each spectrum had S/N $\sim 5$.  The number of ``young" spectra relative to ``old" spectra in the parent populations was increased in steps of 5: the first parent population 25 noise realizations of young galaxy; the second parent population contains 5 noise realizations of the old galaxy and 20 noise realizations of the young galaxy; the third has 10 noise realizations of the old galaxy and 15 noise realizations of the young galaxy; and so forth, with the final parent population containing 25 noise realizations of the old galaxy. For each parent population, we first co-added the galaxies in each sample (for example, for the first parent population, we co-added 25 synthetic 2 Gyr old spectra, each with a different random noise realization; for the second, we co-added 20 synthetic 2 Gyr old spectra and 5 synthetic 7 Gyr old spectra, each with different noise realizations, etc.). We then measured the nominal light-weighted age of each co-added spectrum, marked by the diamonds in Figure \ref{fig:synthhisto}.

Next, on each parent sample, we used the bootstrap resampling technique, which proceeded as follows:  we drew 25 galaxy spectra, with replacement, at random from the sample.  Note that for the parent samples with entirely young or entirely old populations, this effect is reduced to drawing galaxies with different added noise rather than different ages. These randomly selected galaxies were co-added and an age was measured from the resultant spectrum.  This process of drawing at random with replacement from the parent sample, co-adding the drawn galaxies, and measuring an age was repeated 4000 times, resulting in 4000 different age measurements for each parent population.  

The age histograms for the six parent populations with differing percentages of 7 Gyr old galaxies and 2 Gyr old galaxies can be seen in Figure \ref{fig:synthhisto}.  Given that Lick indices have a non-linear response to age, we plot our histograms in log-age, with a bin size of 0.02 in log(age(Gyr)). We note that when we have more homogeneous populations (the young population with 0\% old galaxies and the old population with 100\% old galaxies), we recover a more Gaussian distribution of ages centered roughly around the age of the input spectra.  In the mixed populations, we see tails develop particularly towards older ages.

\begin{figure*}[hbt!]
\includegraphics[trim= 0mm 0mm 0mm 0mm, angle=0, scale=1.0]{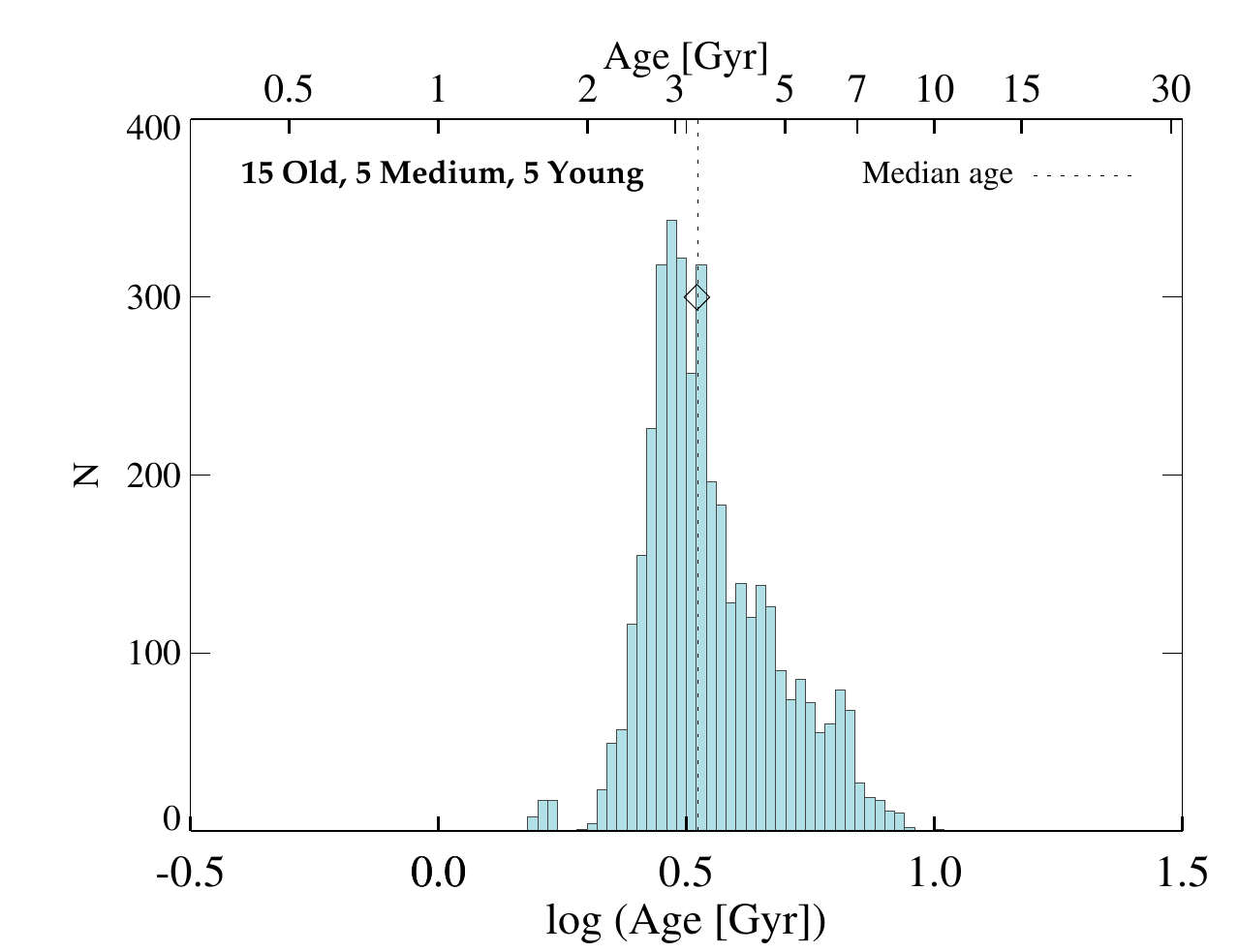}
\caption{Age histograms generated from 4000 bootstrap resamplings of the synthetic galaxy spectra drawn from a parent population with 15 galaxies 7 Gyr in age, 5 galaxies 5 Gyr in age, and 5 galaxies 2 Gyr in age. The histogram has bin size 0.02 in log(age (Gyr)). The median log age is marked by a dotted line. The black diamond denotes the nominal light-weighted age of the original co-added spectrum.}
\label{fig:3age}
\end{figure*}

To explore how far we could recover a mixed population, we created an additional parent population with 25 total galaxies:  15 galaxies of 7 Gyr age, and 5 galaxies each with 5 Gyr and 2 Gyr age.  This ``mixed" parent population would, potentially, display a three-peaked histogram.  Figure \ref{fig:3age} shows the age histogram for the ``mixed" parent population.  We do not clearly recover three distinct peaks, but the population is distinctively less Gaussian than our homogeneous populations. It closely resembles the synthetic population with 15 galaxies of 7 Gyr age and 10 galaxies of 2 Gyr age, which is unsurprising. 

We note that another informative test of this procedure would be to perform the bootstrap resampling a number of times on synthetic spectra with random age distributions instead of two distinct ages. However, we performed this initial simpler test to determine if the procedure could pick out distinct populations if they existed. These simple synthetic galaxy comparisons have shown that mixed populations do indeed leave an imprint in the bootstrap-resampled age histograms, but that it is difficult to tell precisely what the degree of heterogeneity is within the stellar population ages.  Our approach is to therefore exploit this information to characterize the homogeneity of galaxy populations. 

\section{Results}
\label{sec:results}

In this section we apply the bootstrapping technique to the samples of early-type galaxies defined in \S\ref{sec:samples}. We drew galaxies at random with replacement from a given subsample. The total number of galaxies drawn is equal to the size of the given subsample; for example, for the RBD Compact subsample, we draw 25 galaxies each time, but for the RBD Control sample we draw 53 galaxies. We co-added the randomly drawn galaxies, and then measured the resultant age with EZ\_Ages. This procedure was repeated 4000 times for each subsample of galaxies, giving nearly 4000 estimates for each sample's age.\footnote{Note that due to the time-intensive process of measuring the velocity dispersions using the IRAF cross-correlation technique, we did not measure the velocity dispersion of each new co-added spectrum in our bootstrap.  Instead, we performed a number of small (300 re-samplings) bootstrap co-additions, measuring both the velocity dispersion and age for each re-sampling.  We then performed the same analysis with 300 re-samplings, but this time assuming the velocity dispersion for each co-add was the same as the velocity dispersion measured in the original sample. We found negligible difference between these two bootstrap distributions.  Therefore for our large bootstrapping procedure, the velocity dispersion for each of the new co-added spectra was assumed to be the same as the measured velocity dispersion of the original sample from which we performed the resampling.}

\begin{figure*}[hbt!]
\includegraphics[trim= 0mm 0mm 0mm 220mm, angle=0, scale=2.0]{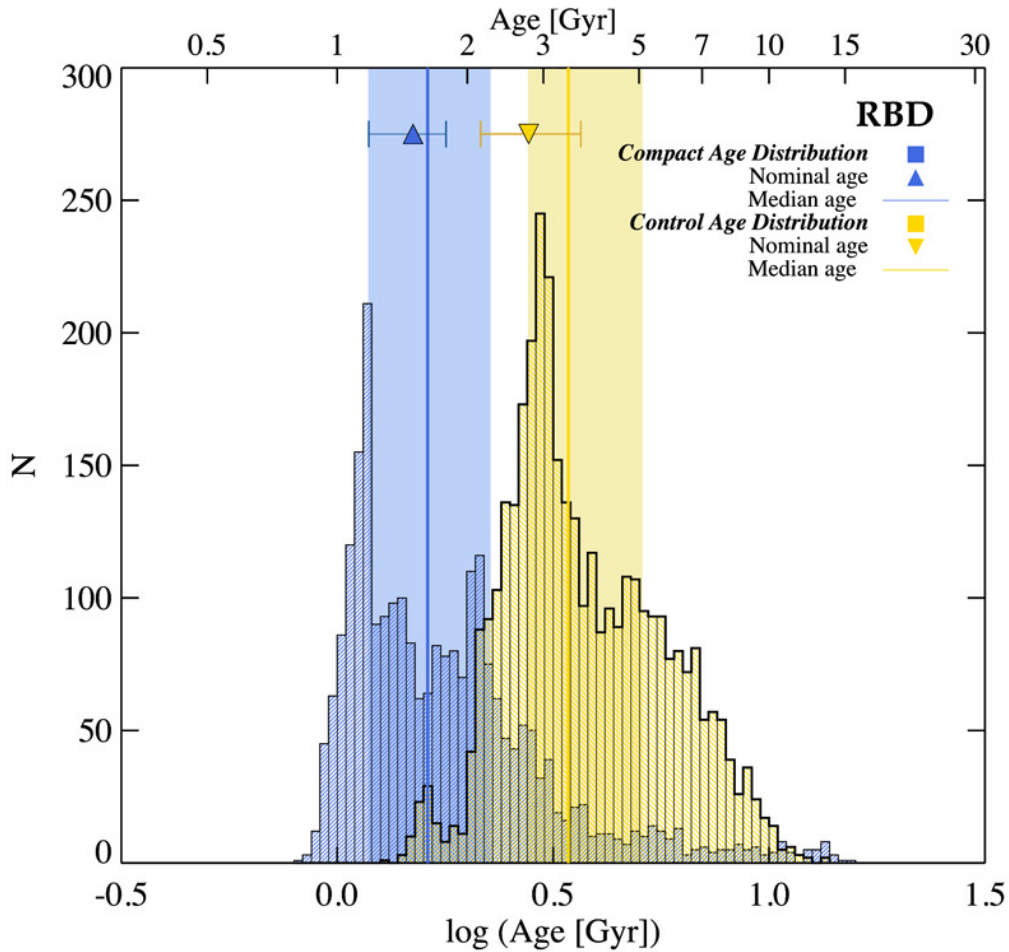}
\caption{Age histograms generated from bootstrapping the Compact (blue) and Control (yellow) galaxies in the ``Red \& Bulge-Dominated (RBD) Sample" 4000 times each.  The histograms have bin size 0.02 in log(age (Gyr)).  Measured nominal ages from the original (non-resampled) co-added subsamples are demarcated by symbols of the same colors. The errors are estimated by EZ\_Ages using the signal-to-noise spectrum to determine the measurement error of the Lick index line strengths. The median ages from the bootstrap resamples are marked with a solid line of the same color as in the histogram. The interquartile range, a measure of dispersion which encompasses 25\% of the data points on either side of the median age, is shown by a semi-transparent band in the same color as the histogram it is measured from.}
\label{fig:RBDhist}
\end{figure*}

\begin{figure*}[hbt!]
\includegraphics[trim= 0mm 0mm 0mm 220mm, angle=0,, angle=0, scale=2.0]{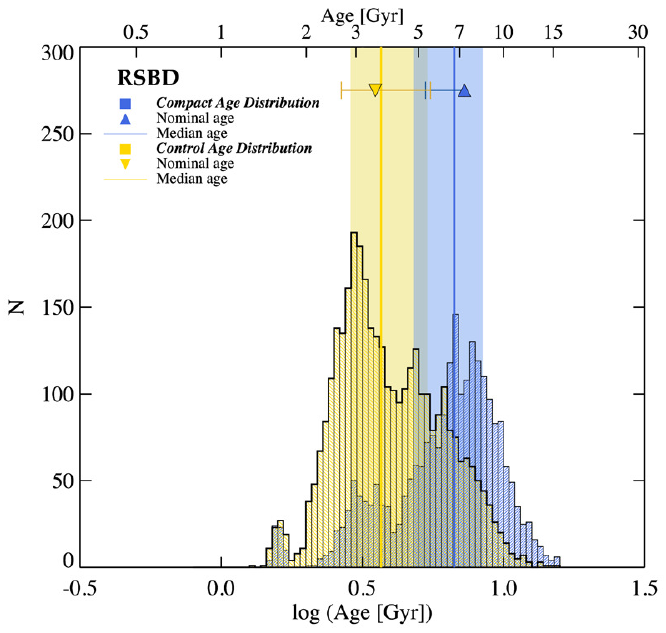}
\caption{Age histograms generated from bootstrapping the Compact (blue) and Control (yellow) galaxies in the ``Red, Smooth, \& Bulge-Dominated (RSBD) Sample" 4000 times each.  The histograms have bin size 0.02 in log(age (Gyr)). Measured nominal ages from the original (non-resampled) co-added subsamples are demarcated by symbols of the same colors. The errors are estimated by EZ\_Ages using the signal-to-noise spectrum to determine the measurement error of the Lick index line strengths.  The positive age limit for the Compact sample cannot be calculated by EZ\_Ages because it exceeds the maximum 15.8 Gyr age of the models. The median ages from the bootstrap resamples are marked with a solid line of the same color as in the histogram. The interquartile range, a measure of dispersion which encompasses 25\% of the data points on either side of the median age, is shown by a semi-transparent band in the same color as the histogram it is measured from.}
\label{fig:RSBDhist}
\end{figure*}

Histograms showing the measured age distributions inferred from the bootstrap resamplings are shown in Figures \ref{fig:RBDhist} and \ref{fig:RSBDhist}.  The age measurements for each of the original co-added samples are shown in Table \ref{table:results}, along with the median and modal ages from the bootstrap resampling.

The central findings emerging from these figures is that massive compact early-type galaxies selected on the basis of red color and high bulge-to-total ratio are {\em younger} than similarly selected larger galaxies, suggesting that size growth in these objects is {\em not} driven mainly by progenitor bias, and that individual galaxies grow as their stellar populations age. However, compact early-type galaxies selected on the basis of image smoothness and high bulge-to-total ratio are older than a control sample of larger galaxies.

\begin{figure*}[hbt!]
\includegraphics[trim= 10mm 0mm 0mm 0mm, angle=0, scale=0.55]{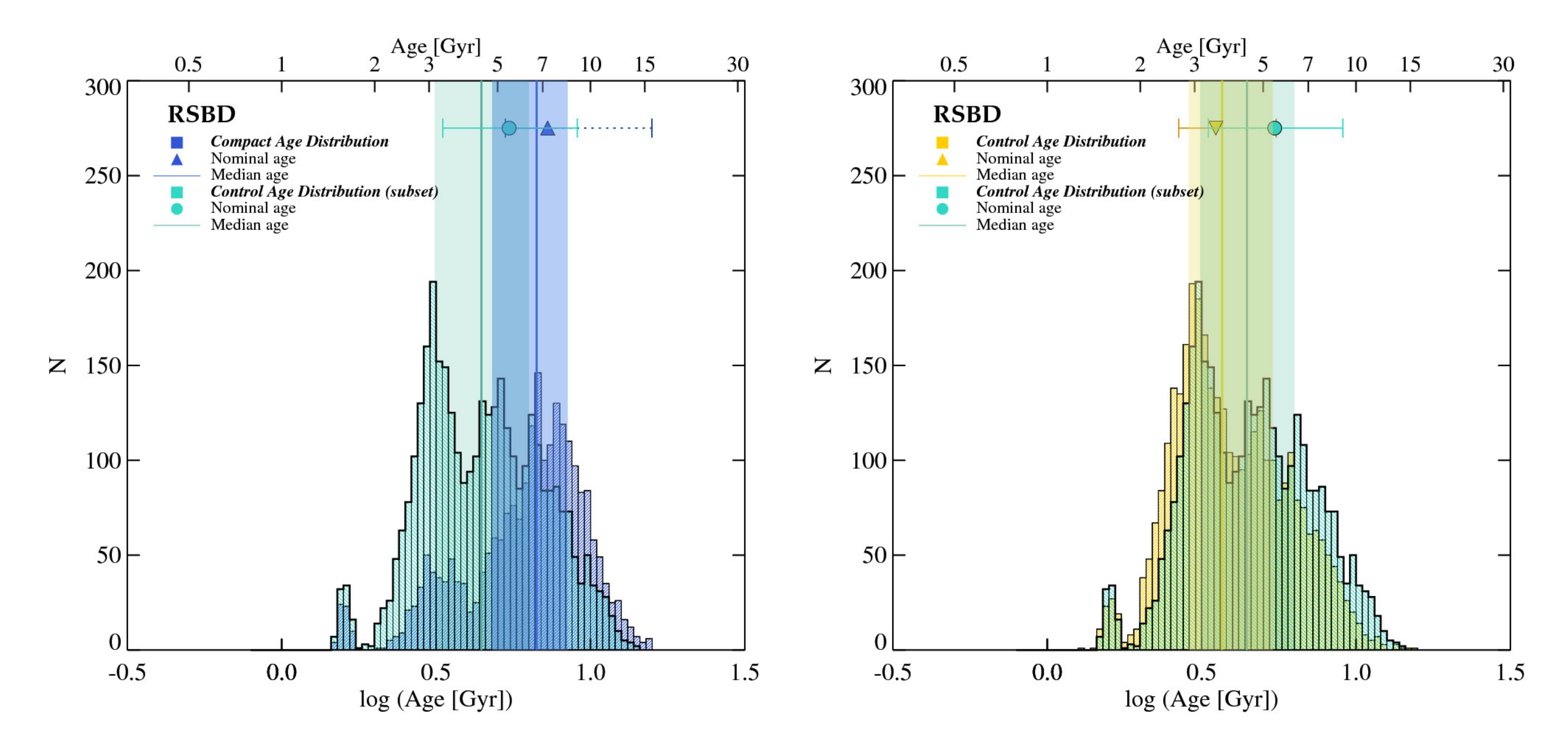}
\caption{Left panel: Age histograms generated from bootstrapping the Compact (blue) and a redshift-matched subset of the Control (teal) galaxies in the ``Red, Smooth, \& Bulge-Dominated (RSBD) Sample" 4000 times each.  The histograms have bin size 0.02 in log(age (Gyr)). Measured nominal ages from the original (non-resampled) co-added subsamples are demarcated by symbols of the same colors. The errors are estimated by EZ\_Ages using the signal-to-noise spectrum to determine the measurement error of the Lick index line strengths.  The positive age limit for the Compact sample cannot be calculated by EZ\_Ages because it exceeds the maximum 15.8 Gyr age of the models. The median ages from the bootstrap resamples are marked with a solid line of the same color as in the histogram. The interquartile range, a measure of dispersion which encompasses 25\% of the data points on either side of the median age, is shown by a semi-transparent band in the same color as the histogram it is measured from. Right panel: Comparison of the age histograms generated from bootstrapping the Control galaxies (yellow) and a subset of the Control galaxies (teal) which have been matched in redshift to Compact galaxies in the ``Red, Smooth, \& Bulge-Dominated (RSBD) Sample" 4000 times each. Symbols as described in the previous panel.}
\label{fig:RSBDzmatch}
\end{figure*}

The basis for these results, organized by parent samples, is as follows:

\begin{enumerate}
\item{\em Red \& Bulge-Dominated}: \\
The results for this sample (selected on the basis of high bulge-to-total ratio and red color) are shown in Figure \ref{fig:RBDhist}.  
In this subsample, we measure (from the original co-added spectra) a nominal age of 1.5 Gyr for the compact galaxies and a nominal age of 2.78 Gyr for the control galaxies.  The median/modal ages measured from the 4000 bootstrap resamplings are 1.62/1.12 and 3.43/2.98 Gyr respectively.  
The age histogram of the control galaxies has a shape that is very similar to our synthetic population with 60\% 7 Gyr old galaxies and 40\% 2 Gyr old galaxies or that of the distribution with three different ages. Such similarities are, of course, not conclusive - any number of other combinations of galaxy ages could produce similarly shaped distributions. It is clear, however, that the distribution is not Gaussian.
The compact galaxies have an age histogram that looks unlike any of our synthetic distributions. It has two distinct peaks: a primary peak at just slightly $> 1$ Gyr, and a secondary peak at slightly $> 2$ Gyr. The peaks are followed by an extensive tail toward older ages. Given that our synthetic distributions were created by sampling galaxies with only two age variations, we conclude that the Compact sample, with a histogram that displays such a long tail towards older ages and thus is unlike any of our synthetic distributions, is likely to have be composed of a population with a broader range of ages than the Control sample.

We note a possibility of contamination from AGNs, as the RBD Compact sample displays some [OIII]-5007 emission, which is characteristic of AGNs with young stellar populations.  We did not apply explicitly apply a correction for emission infill to our bootstrapped spectra, as the results of \S \ref{sec:infill} imply that the impact of emission infill on our age measurements is minimal. Furthermore, the presence of such infill would skew the results towards older ages, not younger.

\item{\em Red, Smooth \& Bulge-Dominated}:\\
The results for this sample (selected on the basis of high bulge-to-total ratio, red color, and smooth morphology) are shown in Figure \ref{fig:RSBDhist}.  
The co-added spectrum of Compact sample has a nominal age of 7.27 Gyr, and the median/modal age of the histogram of ages from the bootstrap resampling is 6.69/6.80 Gyr.  
The Control sample has a nominal age of 3.51 Gyr, over 3.75 Gyr younger than the Compact sample, and the median/modal age of the bootstrap resampling histogram is 3.68/2.97 Gyr. It is tempting to attribute the age difference between the Compact and Control samples to the fact that the two samples have statistically different redshift distributions before the co-addition was performed. However, Figure \ref{fig:c3spec} makes the important point that at the wavelengths of most key spectral features (e.g. H$\beta$ and $\langle$Fe$\rangle$) the mean redshift of the co-added galaxies is very similar for both samples, so that once the galaxies have been co-added (after appropriate normalization) the initial differences in the redshift distributions are not very meaningful. Nevertheless, as a sanity check, we performed the bootstrap resampling on a subset of the Control sample that was more matched in redshift to the Compact sample. To do this, we reduced the number of galaxies at $z > 0.8$ in the Control sample such that the two distributions are the same from $0.8 < z < 1.3$. The results of bootstrapping this new subset can be seen in Figure \ref{fig:RSBDzmatch}. The nominal age of the redshift-matched Control sample was 5.45 Gyr, or 1.9 Gyr older than the nominal age of the original Control sample. The median age of the bootstrap resample did not change as drastically, increasing only by 0.75 Gyr. The shape of the histogram was not significantly altered, either. We therefore conclude that the redshift differences are responsible for some, but not all, of the age differences seen in this sample.

Although the nominal age for the RSBD Control sample is $\sim 0.7$ Gyr older than the nominal age for the RBD Control sample, the bootstrap resampling age histograms for the control samples are markedly similar. The histogram of the control sample has a tiny peak at $\sim 1.5$ Gyr, and then a strong peak at its modal age, with a declining tail to older ages. This, again, looks much like our synthetic population with 60\% 7 Gyr old galaxies and 40\% 2 Gyr old galaxies.  
The Compact sample has a prominent peak at 6.8 Gyr and displays a smaller, secondary peak at the $\sim 3$ Gyr mark, and a third, tiny peak at $\sim 1.5$ Gyr. The Compact sample does not share a distribution shape that is distinctly similar to any of the synthetic populations: although the synthetic distribution with 40\% old galaxies displays three peaks, its peaks are located at different ages and have differing heights. We believe this implies that the stellar populations of the compact galaxies are less homogeneous than that of the control galaxies, but this conjecture remains mostly at a qualitative level.

\end{enumerate}

A two-sample K-S test indicates that there is $<1\%$ likelihood (for both the RBD and the RSBD samples) that the null hypothesis (which is that the compact and control subsamples come from the same population) is correct.  However, because there is an overlap between the interquartile ranges of the RSBD samples, we are cautious in interpreting the results as being indicative of significant age differences in the populations. Finally, we also note that the range of recovered ages extends beyond the age of the universe over the redshift range of the sample, as the age of the universe is $\sim 8.6$ Gyr at the lowest redshift of 0.5.

The main point that emerges from Figures \ref{fig:RBDhist} and \ref{fig:RSBDhist} is the fact that whether the compact galaxies are older or younger than the control galaxies depends on the method  used to define that sample of early-type galaxies.  This result echoes that of \cite{Moresco2013}, who reached similar conclusions coming from a completely different direction, using photometric data from zCOSMOS 20-k sample. 

\begin{deluxetable}{llcccc} 
\tablewidth{0pc}
\tabletypesize{\small}
\tablecaption{Details of Cuts Made for Each Sample, Along with Results}

\tablehead{
\colhead{Sample}		&	\colhead{B/T}			&	\colhead{S2} 		& \colhead{Measured}	& \colhead{Median Boot}	& \colhead{Modal Boot}\\
\colhead{Name}		&	\colhead{cut}			&	\colhead{cut} 		& \colhead{Age (Gyr)}	& \colhead{Age (Gyr)}	& \colhead{Age (Gyr)}\\
} 
\startdata
RBD, Compact		&	$>0.5$	&	N/A			& 1.50	&	1.62		&	1.12\\
RBD, Control		&	$>0.5$	&	N/A			& 2.78	&	3.43		&	2.98	\\
RSBD, Compact	&	$>0.35$	&	$<0.075$		& 7.27	&	6.69		&	6.80\\
RSBD, Control		&	$>0.35$	&	$<0.075$		& 3.51	&	3.68		&	2.97\\

\enddata
\tablecomments{`Compact' and `control' denote radius cuts of $r < 2$ kpc and $r > 2$ kpc, respectively.  Both RBD and RSBD samples have mass $M > 10^{11} M_{\Sun}$ and color $U-B >0.9$.}
\label{table:results}
\end{deluxetable}


\section{Discussion}
\label{sec:discussion}

\subsection{Correlation between ``Smoothness" and Age}
\label{sec:correlation}

When we compare the measured ages of our samples defined without a measure of image smoothness (RBD sample) to those selected with the smoothness criterion (RSBD sample), the ages of both the Compact and Control subsamples of the RBD sample are younger than the ages of both the Compact and Control RSBD subsamples, although the ages of the RBD and RSBD Control samples are consistent with each other within a 1-$\sigma$ uncertainty. However, the significant age difference between the Compact samples reveals a correlation between the ``smoothness" and age at least for compact galaxies: smoother galaxies are older than clumpier ones.

Such a result is not unexpected, as the RSBD sample was particularly chosen to minimize contamination from S0/Sa type galaxies; as such, we considered the possibility that such galaxies are contaminating the population selected in the RBD sample. We visually inspected images of the galaxies in both of our samples in order to discern whether we had such contamination. The following fractions of galaxies were identified to be probably spiral or lenticular galaxies: $\frac{2}{25}$ (8\%) in the RBD Compact sample, $\frac{6}{53}$ (11.3\%) in the RBD Control sample; $\frac{1}{25}$ (4\%) in the RSBD Compact sample, and $\frac{2}{44}$ (4.5\%) in the RSBD Control sample. Images and spectra for these galaxies are attached in the Appendix. The higher percentage of galaxies identified as non-ellipticals in the RBD sample confirms that there is a degree of contamination that is less present in the RSBD sample, and at least partially explains the younger overall ages of the RBD subsamples. The RBD Compact galaxies tend to have a slightly higher redshift than the RSBD Compact galaxies as seen in Figure 2, which may partially explain their older ages. A K-S test between the two distributions indicates, however, that they are not statistically different. Furthermore, the redshift for both samples declines in a similar way within the co-added spectra, as seen in Figures 3 and 4. We therefore conclude that this is unlikely to explain much of the age difference.

In effect, our results are consistent with the simple idea that  adding a disk component to a galaxy decreases its smoothness, and since the disk is likely to be younger than the bulge of the galaxy, adding a disk also lowers the galaxy's mean age. However, it is harder to explain why the effect appears to be differential in nature, with the ages of larger early-type galaxies being relatively insensitive to smoothness. The addition of a small disk might be expected to make a bigger difference to the observed size and clumpiness of a compact galaxy than it would to a relatively large galaxy, which might explain at least part of this effect. In any case, higher resolution observations of compact galaxies that clearly show the existence of disks in these systems and allow their sizes to be measured as part of a multi-component model would allow  these ideas to be tested.

\subsection{Defining an ``Early-type Galaxy" and the Influence on ``Progenitor Bias"}
\label{sec:precision}

It is clear from the differing age measurements obtained in our samples that investigations of stellar populations at high redshifts must be very careful in their definitions of what is meant by an `early-type galaxy' in order to avoid bias. Morphology matters, in addition to color. Yet few of the studies mentioned earlier in \S \ref{sec:context} employ a measure of image smoothness in their sample selection. Our RBD sample is closer than our RSBD sample to the selection generally used by investigations which discuss the size evolution of massive galaxies, as most of the studies employ some measure of morphology (but rarely smoothness) and/or color in selecting their samples.

For example, \cite{Chevance2012}, who investigated the structure of compact massive quiescent galaxies at $z \sim 2$, used the same $B/T > 0.5$ and $S2 \le 0.075$ cuts as our RSBD sample (though without the color cut), but only for selecting their local early-type galaxy sample. For their high-redshift sample, they utilized the color-selected samples of \cite{vanderWel2011} and a variety of surveys compiled by \cite{Damjanov2011} which provides an overview of the selection criteria used by each. Of the 16 spectroscopic surveys examined in \cite{Damjanov2011}, eight are spectroscopically selected objects with old stellar populations \citep{Saglia2010, vanderWel2008, Longhetti2007, Damjanov2011, Damjanov2009, Cimatti2008, Daddi2005, vanDokkum2008}, four are morphologically selected ETGs \citep{Schade1999, Treu2005, Bundy2007, Newman2010, Gargiulo2011, Saracco2011}, and four are quiescent galaxies selected by color \citep{Rettura2010, Ryan2012, Carrasco2010, Cassata2010}.

\cite{Trujillo2011}, who observe that that smaller galaxies (at fixed stellar mass) are \textit{not} older than the larger galaxies, use a sample of visually classified ETGs from the GOODS and SDSS data sets. \cite{Whitaker2012} used color cuts to isolates samples of recently quenched galaxies from the NEWFIRM Medium-Band Survey, and found that younger quiescent galaxies are not larger, and in fact may be somewhat smaller, than older galaxies at a fixed redshift. \cite{HuertasCompany2013} studied morphologically selected quiescent ETGs from the COSMOS survey from $z \sim 1$ to the present and found that galaxy size-mass relation and size growth do not depend on environment.  Most recently, \cite{Morishita2014} select quiescent galaxies from the MOIRCS Deep Survey and \textit{HST}/WFC3 CANDELS data in the GOODS-N region using rest-frame colors and find the size growth for massive quiescent galaxies to be consistent with previous studies, at a factor of $\sim 2.5$ increase from $z \sim 2.5$ to $z \sim 0.5$ at a given stellar mass. \cite{vanderWel2014}, with spectroscopy and photometry from 3D-\textit{HST} and imaging from CANDELS, used rest-frame colors to isolate quiescent galaxies and showed that the number density of small, compact ETGs strongly decreases between $z \sim 1.5$ and the present.

Other studies have concluded that the method of choosing ``early-type" galaxies is important. Our central idea is consistent with \cite{Bernardi2010}, who compared samples selected using photometric and spectroscopic information with those based on morphological information and find that samples selected on the basis of colors alone run the risk of being highly contaminated by edge-on disks, which are the reddest objects at intermediate luminosities or stellar masses. They suggest that the additional requirement of an axis ratio selection $b/a \ge 0.6$ would provide a simple way to select relatively clean early-type samples in high-redshift data sets. We find our results are in strong agreement with \cite{Moresco2013}: they selected six samples of early-type galaxies up to $z=1$ from the zCOSMOS-20k spectroscopic survey and analyzed the samples' photometric, spectroscopic, and morphological properties.  Their samples were based on morphology, optical colours, specific star formation rate, a best-fit to the observed SED, and a criterion that combined morphological, spectroscopic, and photometric information.  They found that the level of contamination from blue, star-forming, or otherwise non-passive outliers was highly dependent on the method by which the sample was selected.  The sample selected by morphological criteria (a combination of principal component analysis of five nonparametric diagnostics of galaxy structure and a parametric description of galaxy light) displayed the highest percentage of contamination and showed significant emission lines in the median stacked spectra.  The sample that displayed the least amount of contamination was the one selected to be ``purely passive" by combining multiple selection criteria using morphology, spectroscopy, and photometry.  They also found a strong dependence of the contamination on stellar mass, and concluded that regardless of the adopted selection criteria, a significantly purer sample can be obtained with a cut at $M > 10^{10.75} M_{\Sun}$. 

As described earlier, massive compact early-type galaxies selected on the basis of red colors and high B/T ratios display younger ages than the control sample of larger galaxies at similar redshifts and in a similar mass bin.  As ``progenitor bias" posits that younger galaxies are larger at fixed mass, we therefore conclude that progenitor bias cannot account for the size growth of compact galaxies, as defined by our RBD selection, and that the individual galaxies experience growth as their stellar populations age.  However, the RBD sample is only one reasonable way to isolate early-type galaxies.  Using other approaches, we arrive at a different conclusion.

Compact galaxies that are selected on the basis of image smoothness and high B/T ratios display older ages than the control sample of larger galaxies, a result that is consistent with the size growth explanation of progenitor bias.  In their recent paper, \cite{Carollo2013} use the large COSMOS survey to argue that progenitor bias can explain most of the observed size growth of compact galaxies, with size changes due to merging and other processes being of secondary importance, particularly for objects with masses below $10^{11} M_{\Sun}$.  We conclude that progenitor bias can indeed play a significant role in explaining the apparent size growth of early-type galaxies, but only if they are selected on the basis of the smoothness of their light distributions.  We conclude that the importance of progenitor bias in driving the growth of galaxies is surprisingly sensitive to these sorts of details. 

We note that if, as \cite{Carollo2013} suggests, many local compact galaxies are indeed simply misclassified or missing from the SDSS due to seeing, this would not be the case with data from the Canada-France-Hawaii Telescope Legacy Survey (CFHTLS). An analysis of the local fraction of compact early-type galaxies in a survey such as CFHTLS would be a useful way to determine whether or not that is the case.

\section{Conclusions}
By exploiting the statistical technique of bootstrap resampling, we have explored a method for characterizing the distribution of stellar populations in co-added spectra and investigated the importance of progenitor bias in explaining the rarity of compact massive galaxies in the local universe. We looked for systematic differences in the stellar populations of compact early-type galaxies in the DEEP2 survey as a function of size by comparing the light-weighted ages of compact  early-type galaxies at redshifts $0.5 < z < 1.4$ to those of a control sample of larger galaxies at similar redshifts and in similar mass bins. All galaxies in our sample are selected with the same red color cut. However, massive compact early-type galaxies selected on the basis of high bulge-to-total ratio are found to be younger than similarly selected larger galaxies, suggesting that size growth in these objects is {\em not} driven mainly by progenitor bias. In this sample, the bulk of the size growth is consistent with individual galaxies growing with time. However, compact early-type galaxies {\em selected on the basis of image smoothness}, in addition to high bulge-to-total ratio, are older than a control sample of larger galaxies.  Progenitor bias could well play a significant role in defining apparent size changes in populations of these objects. An important outcome of our study is therefore the surprising sensitivity of conclusions regarding progenitor bias to the definitions used in selecting early-type galaxy populations. This result echoes that of Moresco et al. (2013), who also found that the properties of high-redshift early-type galaxy populations are highly sensitive to the definitions used in defining the samples.

\section{Acknowledgments}
\label{sec:acknowledgments}
We thank the referee for insightful comments and useful suggestions.

We thank Jarle Brinchmann for undertaking the emission line analysis described in  \S \ref{sec:infill}.

S.K. and R.G.A. thank the Natural Sciences and Engineering Research Council of Canada for financial support of this project.

This study makes use of data from AEGIS, a multiwavelength sky survey conducted with the \textit{Chandra}, \textit{GALEX}, \textit{HST}, Keck, CFHT, MMT, Subaru, Palomar, \textit{Spitzer}, VLA, and other telescopes and supported in part by NSF, NASA, and STFC.  

\appendix

{\bf\Large Appendix: Representative Galaxies and Visually Identified non-ETGs}

In this appendix we present representative images and spectra of each of the four samples. We also present the images of the galaxies that we have visually examined and identified as likely to be non-early-type galaxies. Listed in the bottom left corner of each image is the radius, redshift, and mass of the galaxy.

\section{RBD, Compact: Representative Galaxies}

\includegraphics[scale=0.85, trim=10mm 20mm 5mm 0mm]{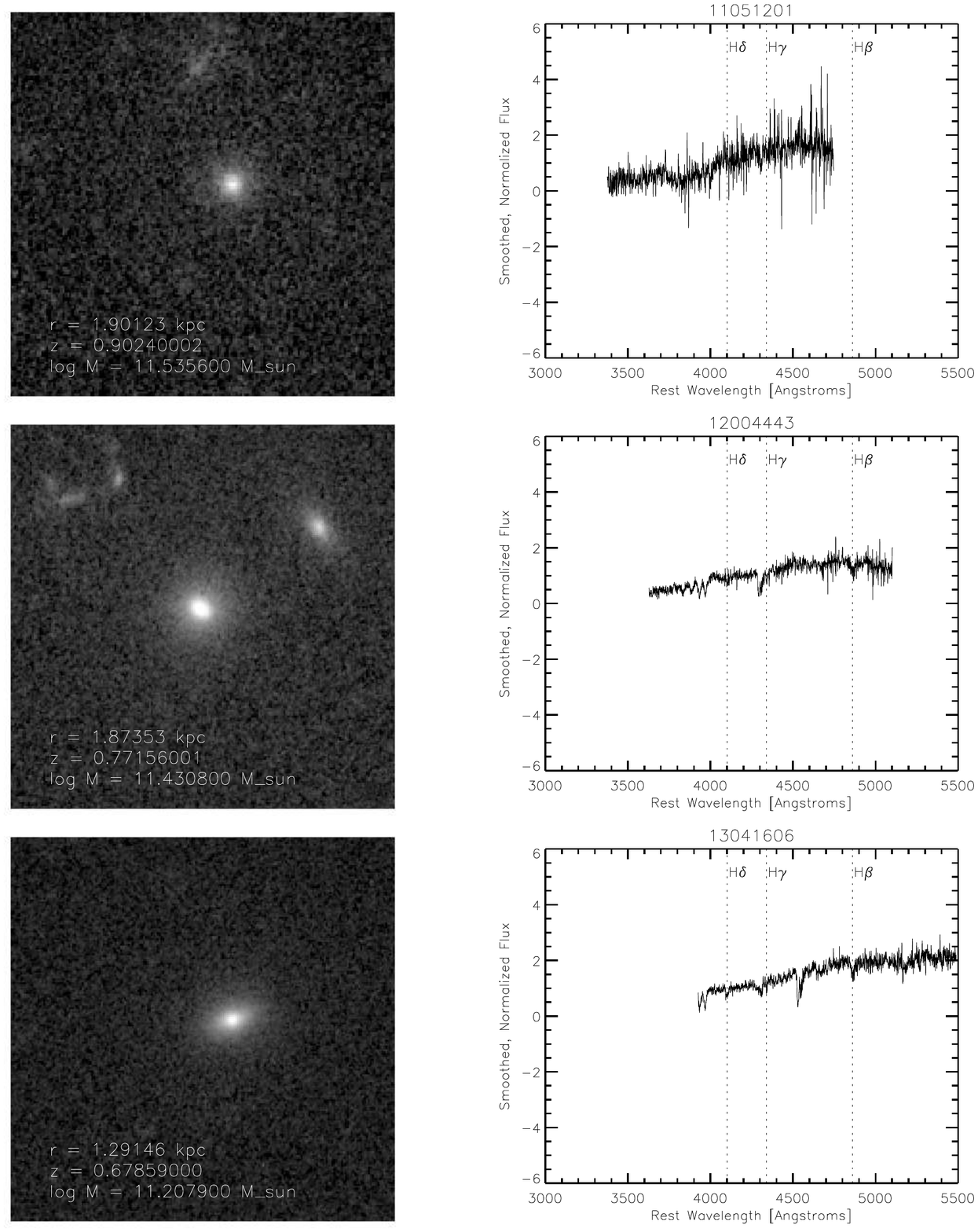}

\section{RBD, Compact: non-ETGs}

\includegraphics[scale=0.85, trim=10mm 20mm 5mm 0mm]{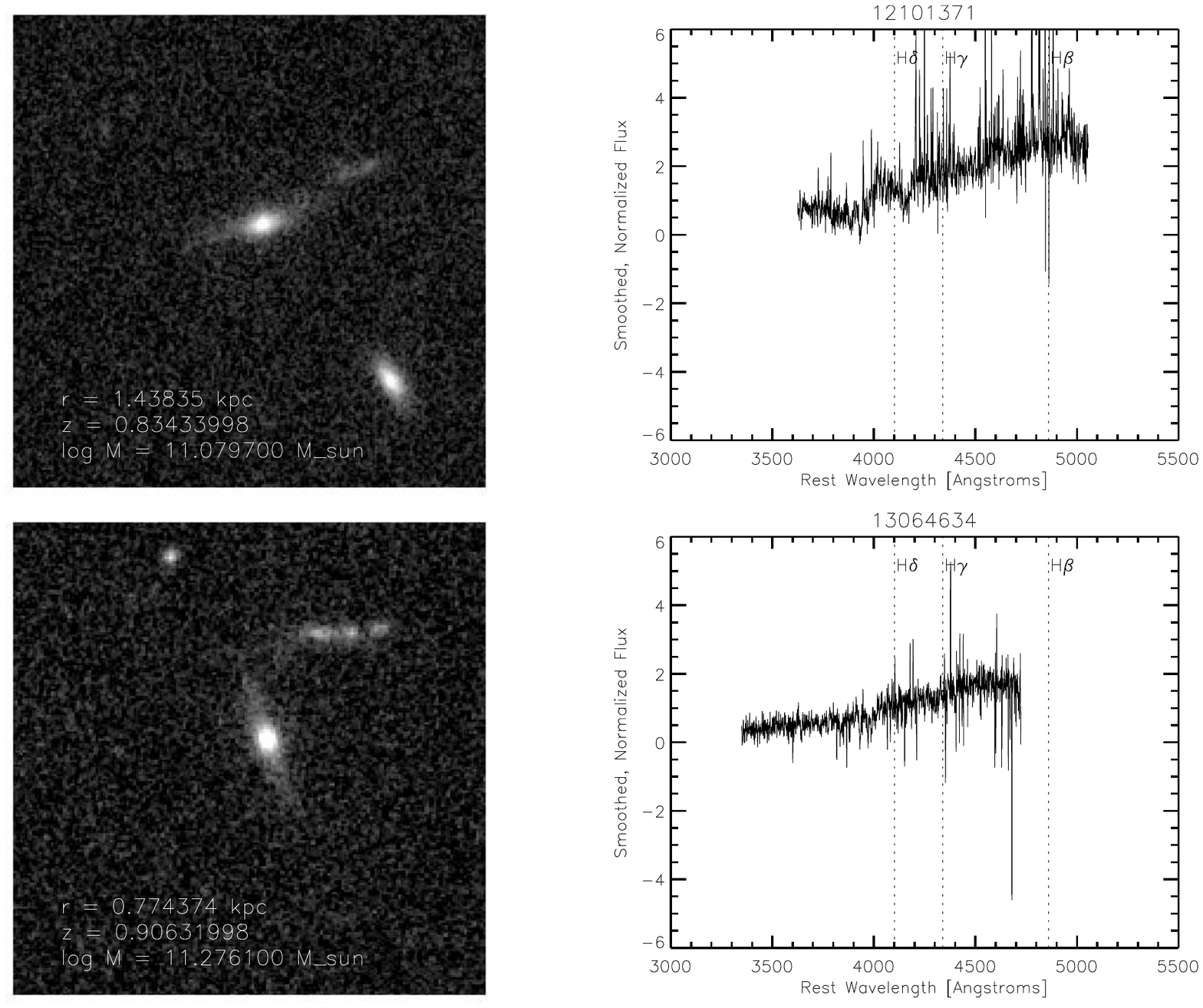}

\section{RBD, Control: Representative Galaxies}

\includegraphics[scale=0.85, trim=10mm 20mm 5mm 0mm]{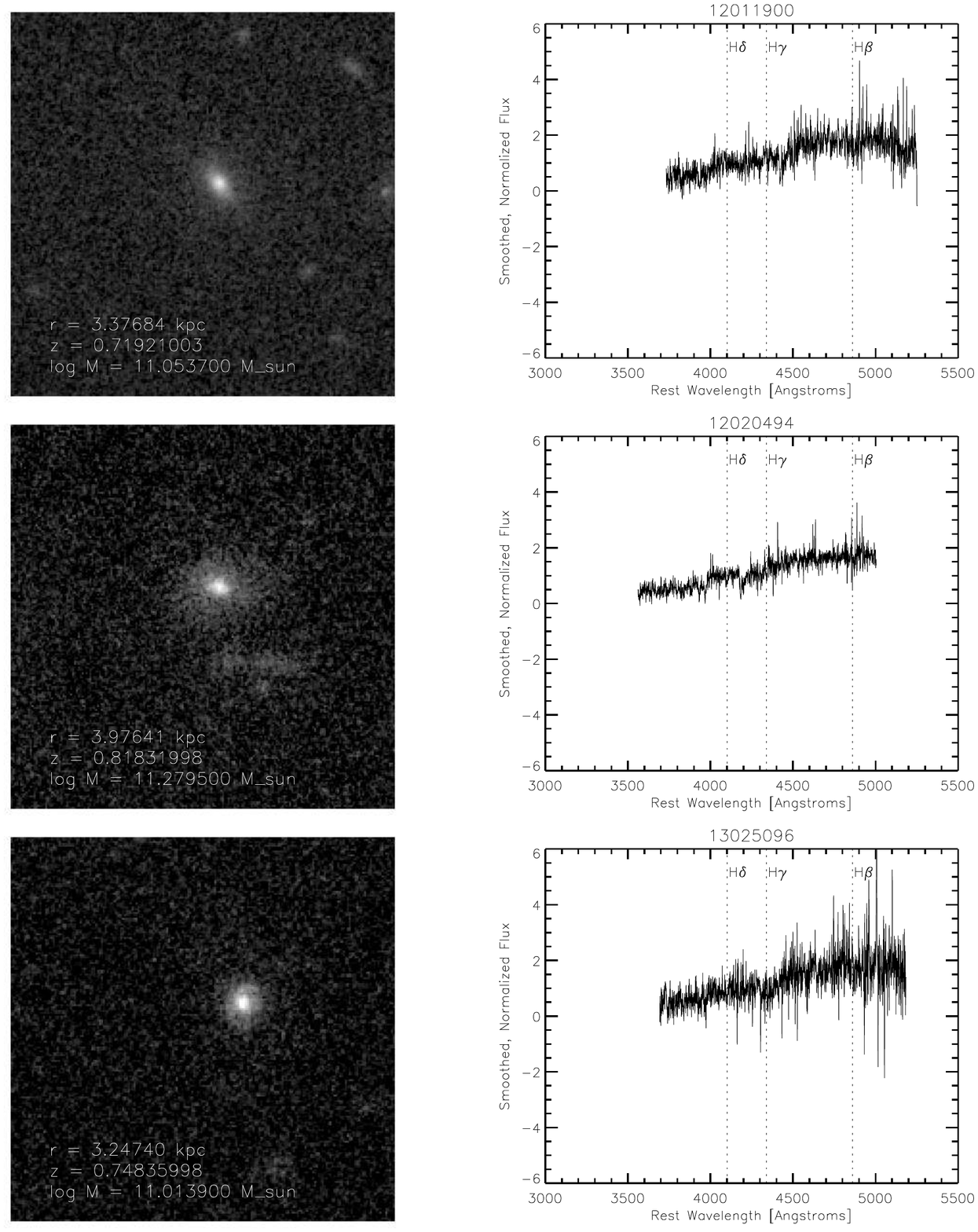}

\section{RBD, Control: non-ETGs}

\includegraphics[scale=0.85, trim=10mm 20mm 5mm 0mm]{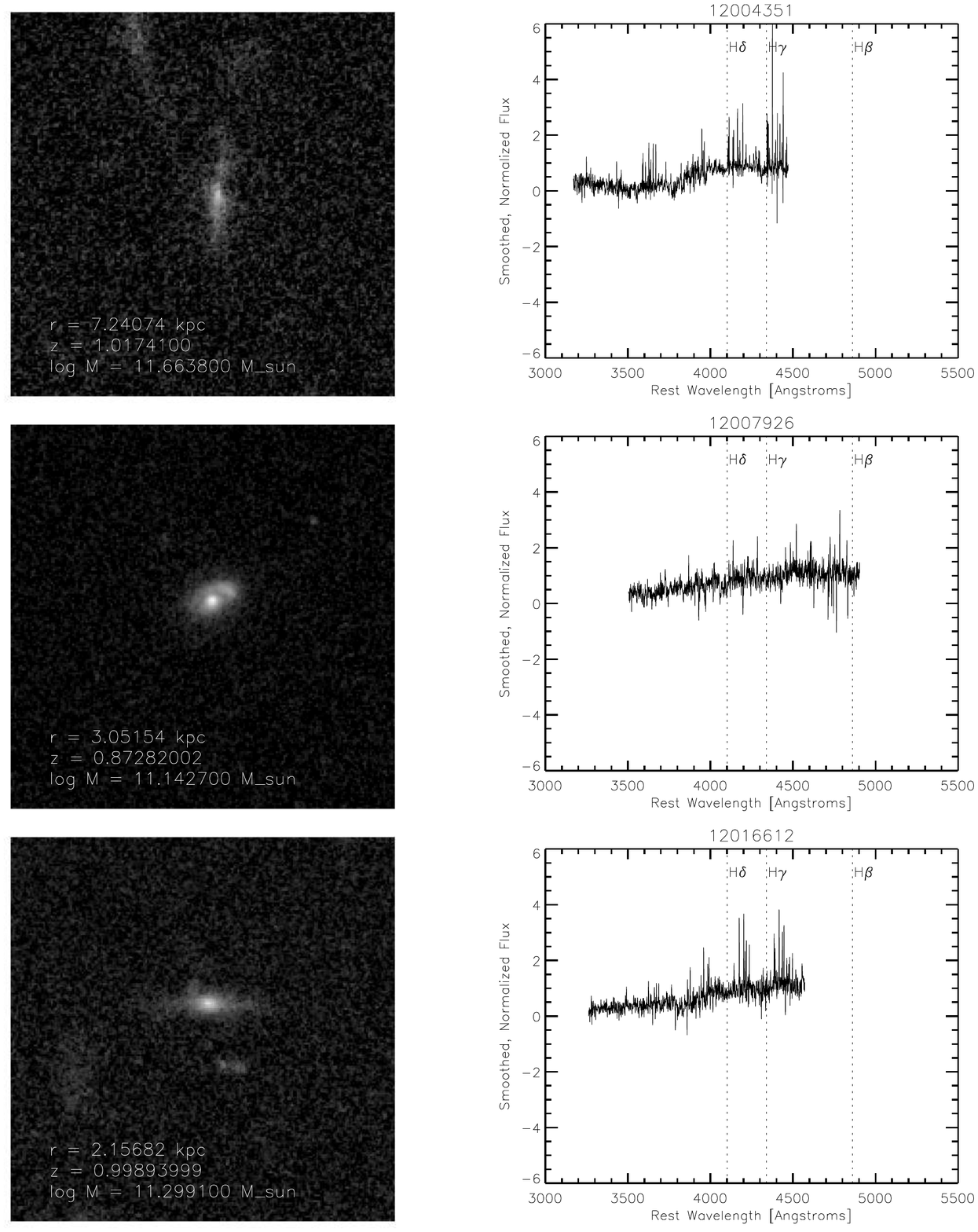}
\includegraphics[scale=0.85, trim=10mm 20mm 5mm 0mm]{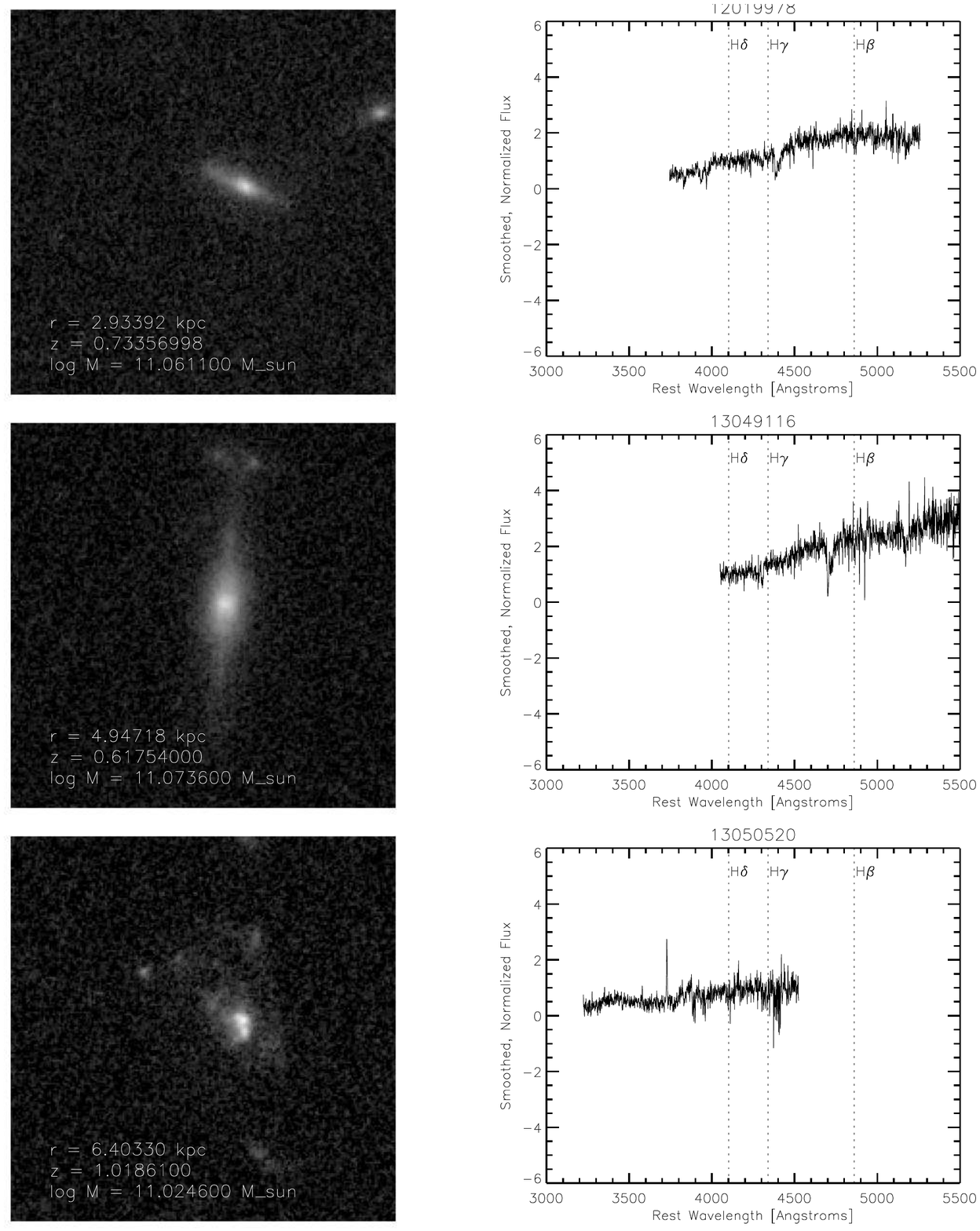}

\section{RSBD, Compact: Representative Galaxies}

\includegraphics[scale=0.85, trim=10mm 20mm 5mm 0mm]{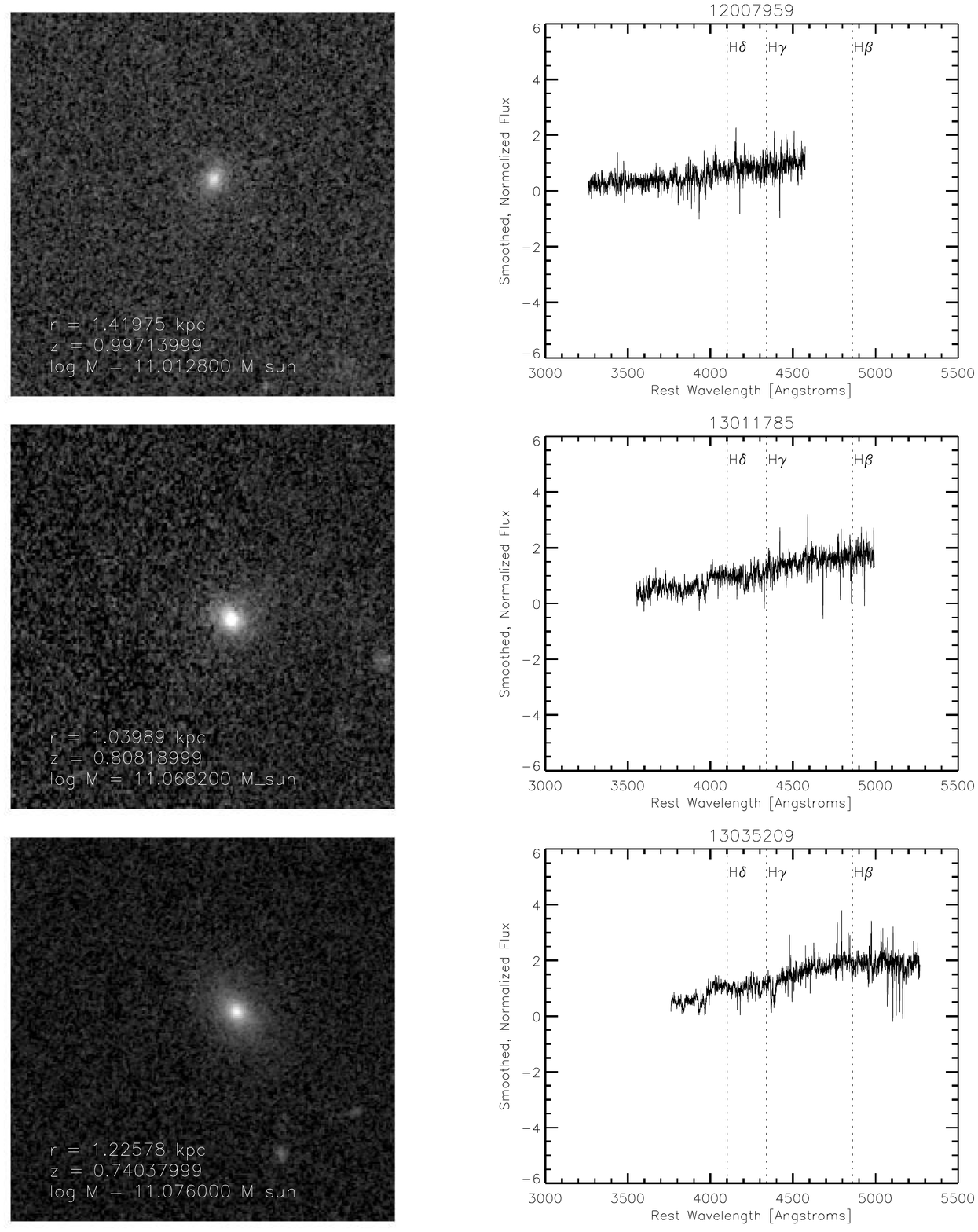}

\section{RSBD, Compact: non-ETGs}

\includegraphics[scale=0.85, trim=10mm 20mm 5mm 0mm]{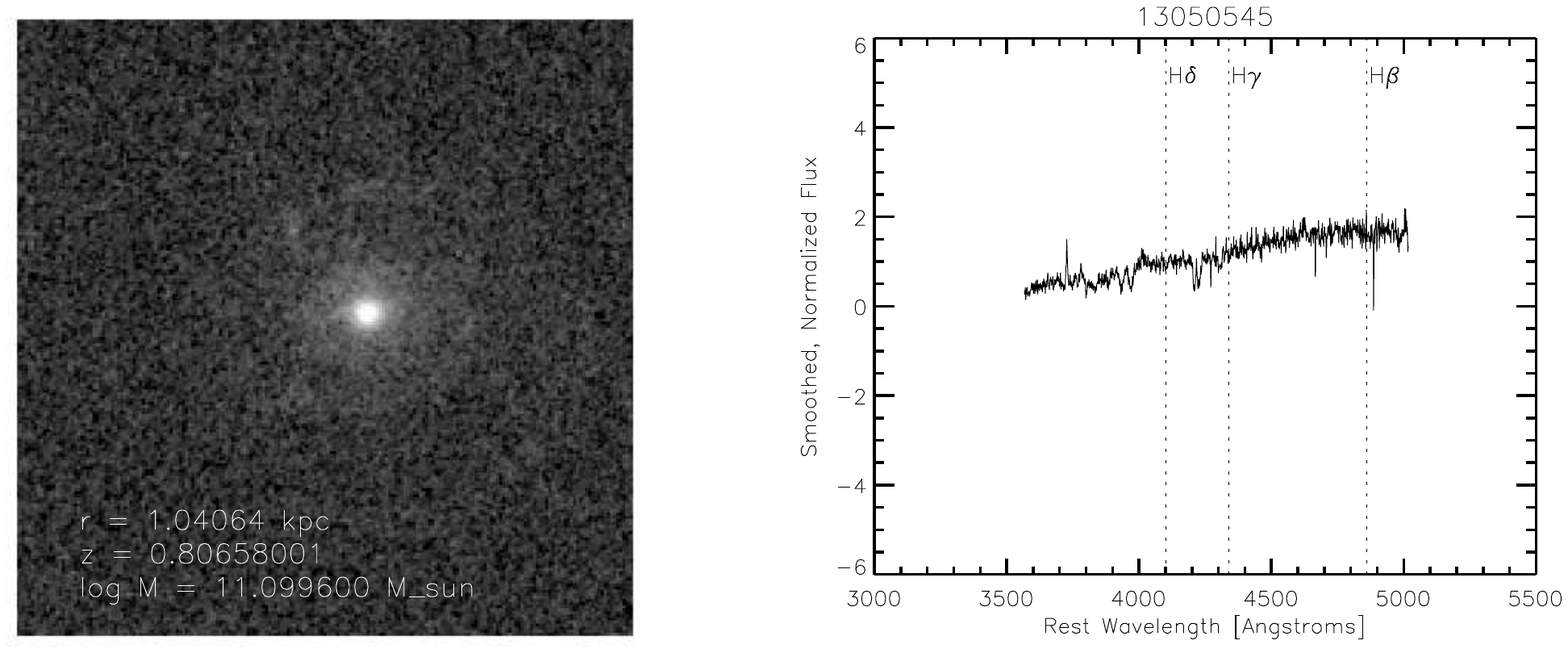}

\section{RSBD, Control: Representative Galaxies}

\includegraphics[scale=0.85, trim=10mm 20mm 5mm 0mm]{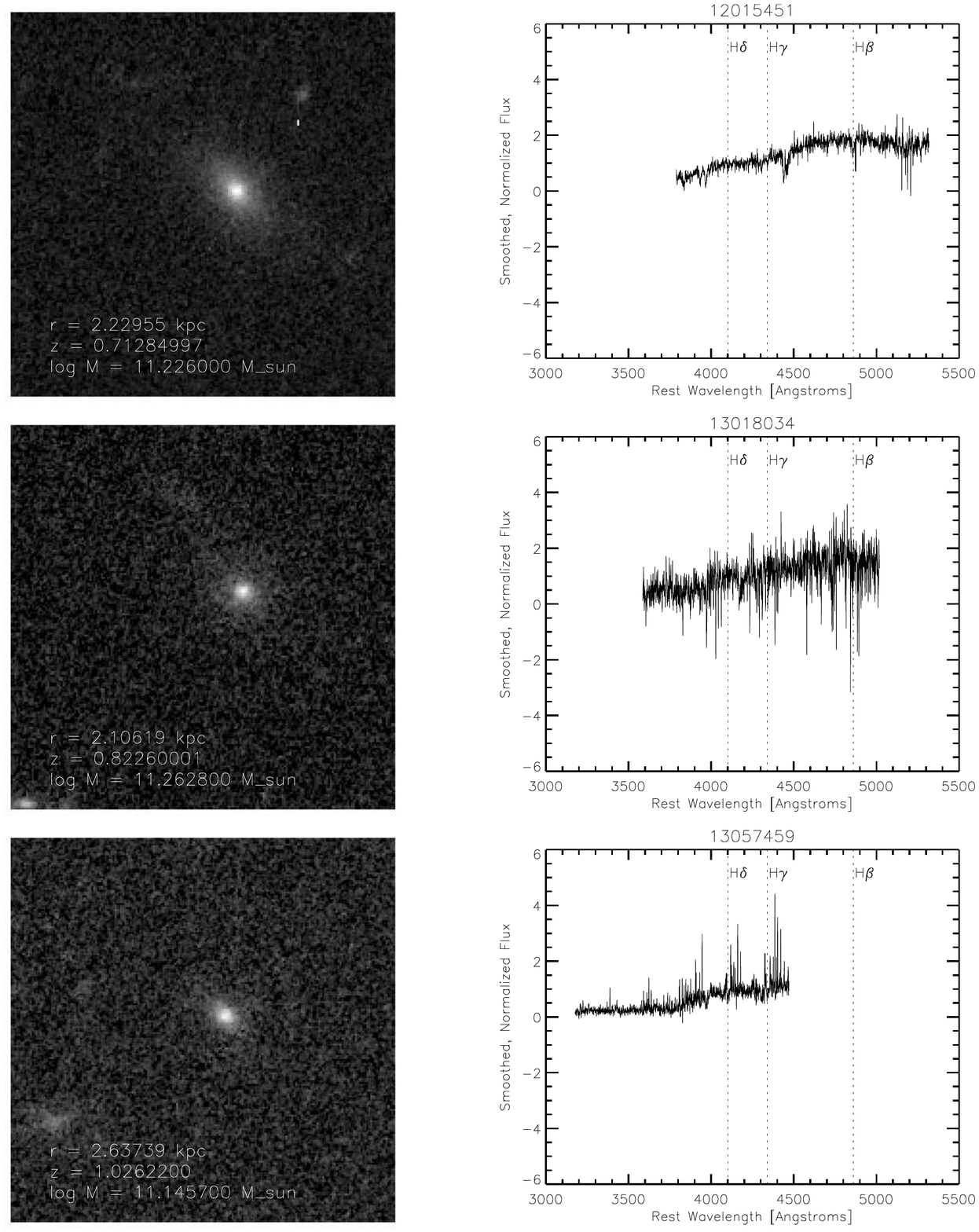}

\section{RSBD, Control: non-ETGs}

\includegraphics[scale=0.85, trim=10mm 20mm 5mm 0mm]{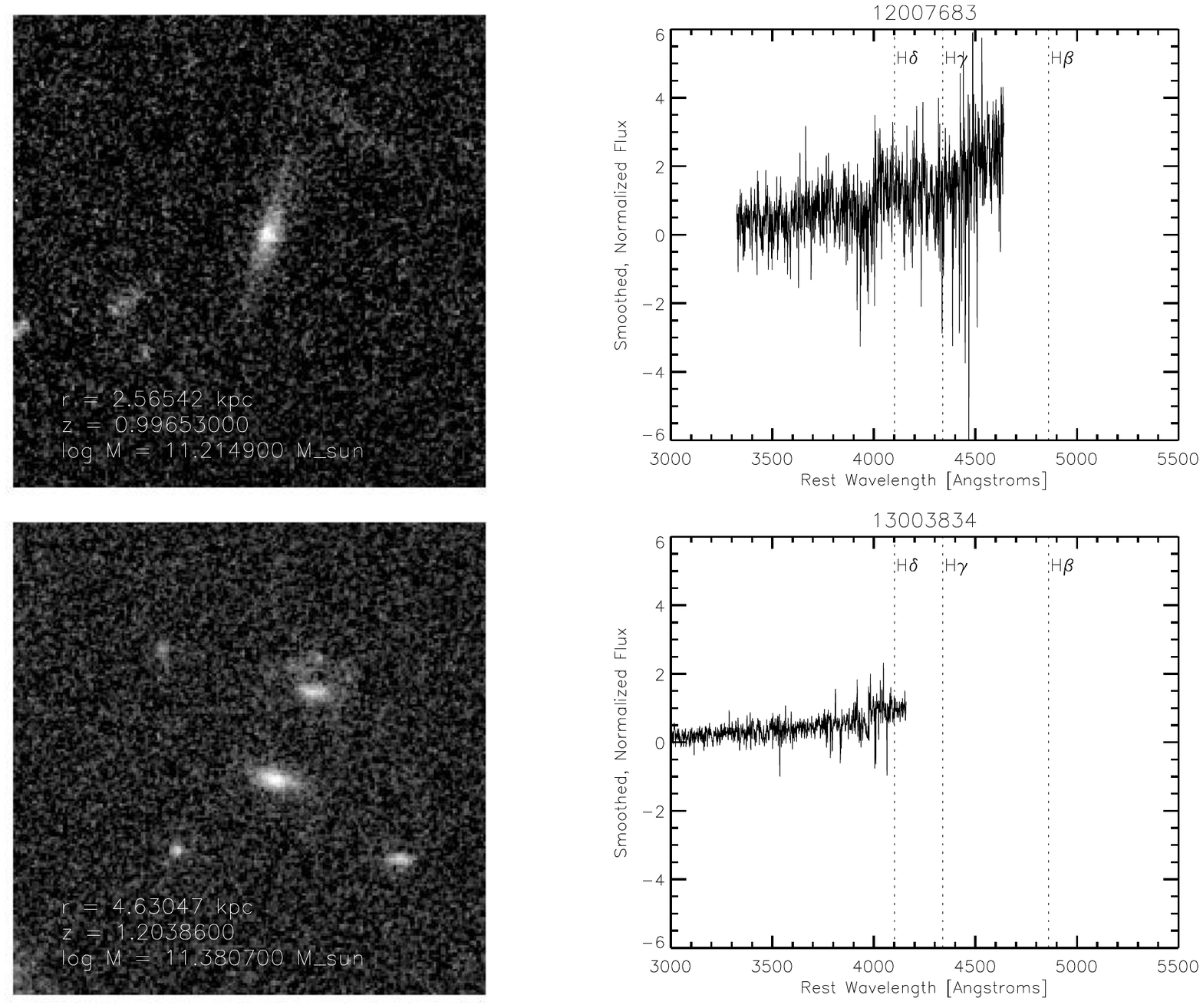}

\clearpage

\bibliography{keating_nov24_2014}

\end{document}